\documentclass[conference]{IEEEtran}
\IEEEoverridecommandlockouts
\usepackage{cite}
\usepackage{color}
\usepackage{array}
\usepackage{comment}
\usepackage{algorithm}
\usepackage[noend]{algpseudocode}
\usepackage{amsmath}
\usepackage{amssymb}
\usepackage{balance}

\usepackage{bm}
\usepackage[pdftex]{graphicx}
\usepackage{caption}
\usepackage[caption=false]{subfig}
\usepackage{url}
\usepackage{pifont}
\usepackage{wasysym}
\newcommand{\protoname}{\textsc{CalTrain}}
\newcommand{\nip}[1]{\vspace{1ex}\noindent\textbf{#1}}

\begin{document}
%don't want date printed
\date{}

%\title{\Large \bf Take \protoname{} to the Confidential and Accountable Land: \\A Collaborative Learning System via Trusted Execution Environments}

\title{Reaching Data Confidentiality and Model Accountability on the CalTrain}

%\author{Anonymous Author(s)}
%for single author (just remove % characters)
%\begin{comment}
\author
{\IEEEauthorblockN{Zhongshu Gu\IEEEauthorrefmark{1}, Hani Jamjoom\IEEEauthorrefmark{1}, Dong Su\IEEEauthorrefmark{1}, Heqing Huang\IEEEauthorrefmark{1},\\\vspace{-1em} Jialong Zhang\IEEEauthorrefmark{3}, Tengfei Ma\IEEEauthorrefmark{1}, Dimitrios Pendarakis\IEEEauthorrefmark{2}, Ian Molloy\IEEEauthorrefmark{1}}\\
\IEEEauthorblockA{\IEEEauthorrefmark{1}IBM Research, \IEEEauthorrefmark{2}IBM Cognitive Systems, \IEEEauthorrefmark{3}ByteDance}
}
%\end{comment}
\maketitle
% Use the following at camera-ready time to suppress page numbers.
% Comment it out when you first submit the paper for review.
%\thispagestyle{empty}

\begin{abstract}
Distributed collaborative learning (DCL) paradigms enable building joint machine learning models from distrusting multi-party participants. Data confidentiality is guaranteed by retaining private training data on each participant's local infrastructure. However, this approach to achieving data confidentiality makes today's DCL designs fundamentally vulnerable to data poisoning and backdoor attacks. It also limits DCL's model accountability, which is key to backtracking the responsible ``bad'' training data instances/contributors. 
In this paper, we introduce \protoname{}\footnote{\protoname{} stands for \underline{C}onfidential and \underline{A}ccountab\underline{l}e \underline{Train}ing System}, a Trusted Execution Environment (TEE) based centralized multi-party collaborative learning system that simultaneously achieves data confidentiality and model accountability. \protoname{} enforces isolated computation on centrally aggregated training data to guarantee data confidentiality. 
%Current TEEs, like Intel's SGX, are performance and capacity constrained. To overcome these limitations, we develop a \emph{partitioned training} strategy that secures the confidentiality of training instances at data entrance, but still benefits from hardware/compiler-level acceleration during model training.
To support building accountable learning models, we securely maintain the links between training instances and their corresponding contributors.   
Our evaluation shows that the models generated from \protoname{} can achieve the same prediction accuracy when compared to the models trained in non-protected environments. 
We also demonstrate that when malicious training participants tend to implant backdoors during model training, \protoname{} can accurately and precisely discover the poisoned and mislabeled training data that lead to the runtime mispredictions.
\end{abstract}
\section{Introduction}
The abundance and diversity of training data are important for
building successful machine learning (ML) models. But,
high-quality training data are scarce. 
Collaborative learning, in which multiple
parties contribute their private data to jointly train an ML model,
aims to address the shortage of high-quality training
resources. However, in many mission-critical and privacy-sensitive
domains, such as health care, finance, and education, training data are tightly controlled by their owners. Sharing raw
data is not permitted by law or regulations.

%To satisfy the requirements of security and privacy from different
%parties, it is crucial to design a privacy-preserving collaborative
%learning mechanism to satisfy the following criteria. Before training
%starts, all training participants (including both training data
%contributors and training infrastructure providers) should make a
%consensus for defining the hyper-parameters for an ML model. During
%the model building process, each individual training participant
%should not be able to view the raw training data or infer the
%information about training data belonging to other data
%contributors. After building the model, the learned model is shared
%among all data contributors.

To meet the security and privacy requirements, multiple distributed
collaborative learning
paradigms~\cite{shokri2015privacy,mcmahan2016communication} have been
proposed to ensure that sensitive training data never leave the
participants' compute infrastructures.  Shokri and
Shmatikov~\cite{shokri2015privacy} proposed a distributed
collaborative training system that exploited the parallelism property
of stochastic gradient descent (SGD). Training participants can
locally and independently build a model with their private datasets,
then selectively share subsets of the model's parameters. In Federated
Learning~\cite{mcmahan2016communication}, a central server can
coordinate an iterative model averaging process. At each training
round, a subset of randomly selected training participants compute the
differences to the global model with their local private training set
and communicate the updates to the central server.

The benefits of client-controlled autonomous data protection come at a
price. These approaches are vulnerable to data poisoning attacks, which can be
instantiated by malicious or compromised training participants. The
reason for this inherent vulnerability stems from how security is
enforced in most distributed learning mechanisms. There, training data
are kept invisible to all participants, except for the data
owner. Consequently, 
%no one, except for the data owner, is able to inspect
%the original data. 
% However, 
malicious data contributors can exploit 
this non-transparency to feed poisoned/mislabeled
training data and implant backdoors into the corresponding
models~\cite{gu2017badnets,liu2018trojaning,
chen2017targeted,bagdasaryan2018backdoor,fung2018mitigating,hayes2018contamination}.
Thus, they can influence and drift the final models'
predictions for their own benefits.

All of the above highlight an important paradox: \emph{data confidentiality is in conflict with model accountability in distributed collaborative learning}. 
Especially with amortized and
stochastic model updates, links
between training data, training participants, and
models have been completely dismantled.  Once model users
encounter erroneous predictions at runtime, they can no longer backtrack
the responsible ``bad'' training data and their provenance.

%In addition, such decentralized collaborative training paradigms also
%face the information leakage problem~\cite{hitaj2017deep,
%melis2018inference} as the dynamically computed model parameters may
%implicitly.

%In addition to decentralized collaborative learning approaches, we
%have also observed a line of research efforts

Separately, there is an emerging trend towards leveraging Trusted
Execution Environments (TEEs), or isolated enclaves, to secure machine
learning training pipelines. For example, Ohrimenko et
al.~\cite{ohrimenko2016oblivious} proposed using Intel Software Guard
Extensions (SGX) to enable multi-party training for
different machine learning methods. More recently,
Chiron~\cite{hunt2018chiron} and Myelin~\cite{hynes2018efficient}
integrated SGX to support private deep learning training services. In
general, current TEE-based training approaches encounter two
performance limiters: (1) TEEs lack hardware acceleration, and (2)
TEEs are memory constrained. As a consequence, it is challenging to
execute deep and complex learning models entirely within an isolated
execution environment.

%To bridge the security and efficiency gap of using TEEs,
%Tram\`er and Boneh~\cite{tramer2018slalom} proposed to securely
%outsource linear layers' computation to out-of-enclave graphics
%processing units (GPUs). Gu et al.~\cite{gu2018securing} advocated a
%vertical layer-wise partitioning strategy to enclose front layers,
%which operate on sensitive inputs, into a secure enclave. But
%currently these two research efforts are only applicable to deep
%learning runtime inference, with no support for training yet.

To address the aforementioned problems, we design and implement \protoname{}, a TEE-based centralized collaborative learning system,
to simultaneously achieve both data
confidentiality and model accountability.
%
%\protoname{} allows training participants to
%provision encrypted training data into a training pipeline. 
\protoname{} uses Intel SGX enclaves on training machines to ensure
the confidentiality and integrity of training data. To
overcome the performance constraints of current SGX, we design a
\emph{partitioned training} mechanism to support learning large-scale deep neural networks with more complex structures. 
% \protoname{} also enforces data authentication and verification to
%deny training data from illegitimate data channels, e.g., data
%injected by training server providers, which are not supposed to feed
%data into the pipeline.
To achieve model accountability, we propose (a) one-way fingerprinting for all training instances and (b) maintaining the links between fingerprints and training
participants.  We ensure that the recorded fingerprints cannot be
reconstructed to reveal the original training data, but can still facilitate debugging incorrect predictions and identifying the
influential training data and their
corresponding contributors.

In our evaluation, we demonstrate that models trained within \protoname{} can effectively protect the confidentiality of private training data and incur no loss on prediction accuracy compared to models trained in regular training environments. 
To verify the effectiveness of model accountability, we test \protoname{} with the \emph{Trojaning Attack}~\cite{liu2018trojaning}, whose authors generously released both their poisoned training datasets and pre-trained models with embedded backdoors to us for reproducing the attack. 
Our experiment shows that we can precisely and accurately identify the poisoned and mislabeled training data, and further discover the malicious training participants.  

To summarize, the major contributions are as follows:
\begin{itemize}
%\vspace{-2ex}
\item \textbf{Confidential Learning:} a TEE-based collaborative
  learning system to protect training data confidentiality,
%\vspace{-2ex}
\item \textbf{Partitioned Training:} a learning workload partitioning
  mechanism to address the performance and capacity constraints of
  existing TEE technologies, and
%\vspace{-2ex}
\item \textbf{Model Accountability:} a data fingerprinting mechanism
  on training instances to support post-hoc provenance and causality
  tracking for mispredictions.
\end{itemize}

%\begin{comment}
\nip{Roadmap.} 
Section~\ref{sec:background} introduces the relevant background knowledge.
Section~\ref{sec:model} talks about the threat model of our proposed
system. Section~\ref{sec:design} details the design principles of
\protoname{}. Section~\ref{sec:implementation} describes the
implementation of our research prototype. Section~\ref{sec:evaluation}
presents the model accuracy, performance, and accountability
experiments as our evaluation. Section~\ref{sec:discussion} discusses
the application scenarios and security implications of potential
attacks. Section~\ref{sec:relate} surveys related work, and we
conclude in Section~\ref{sec:conclusion}.
%\end{comment}

\section{Background}
In this section, we give a brief summary about deep learning,
collaborative training, and Intel SGX.
\label{sec:background}
\begin{figure}[!t]
\centerline{
\includegraphics[width=.45\textwidth]{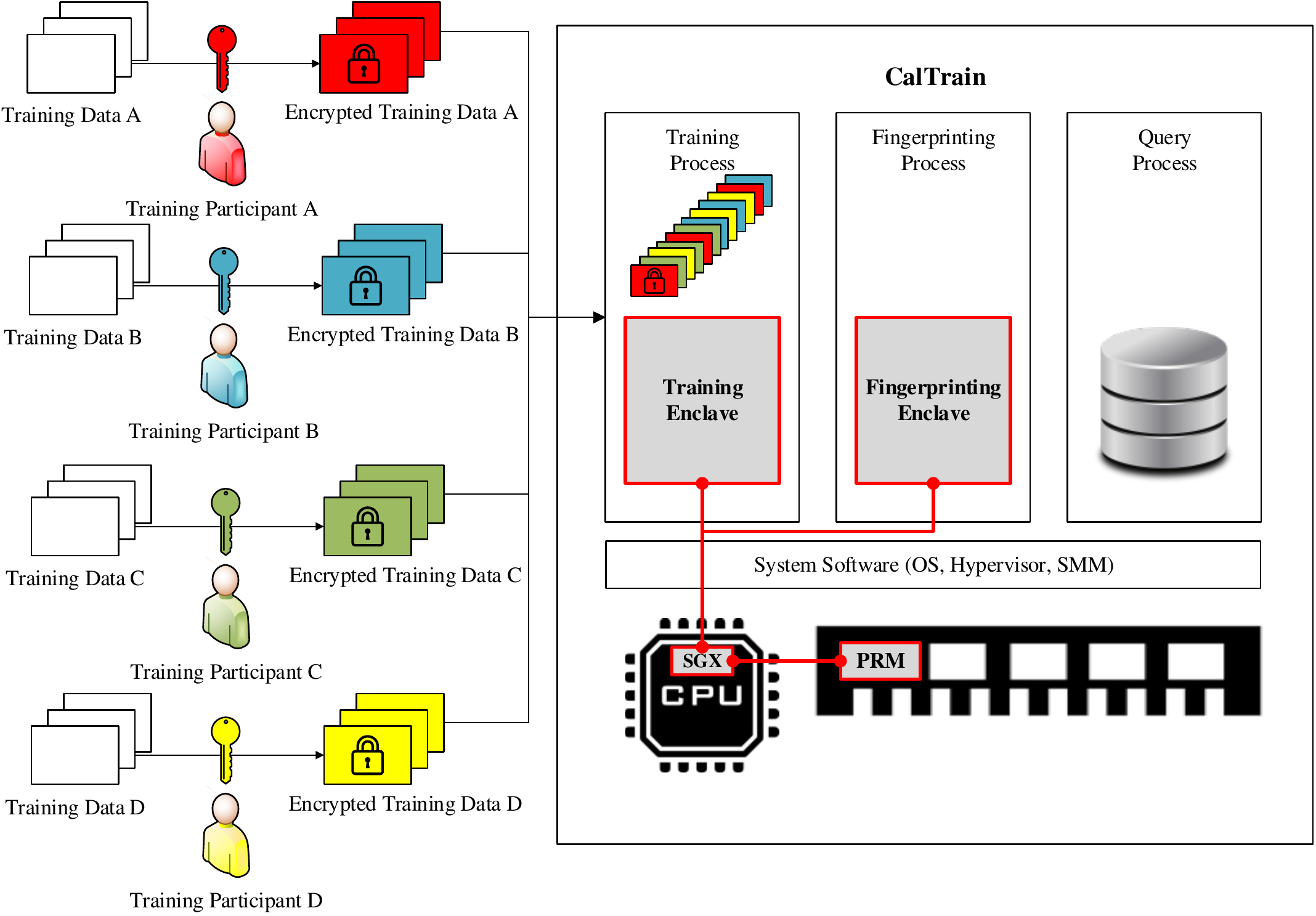}}
\caption{The System Architecture of \protoname{}}
\label{fig:arch}
\end{figure}

\nip{Deep Learning.} Deep learning approaches enable end-to-end
learning by automatically discovering data representations of raw inputs.  
At the core of any deep learning system is a deep
neural network (DNN).  A DNN has multiple hidden layers between the
input and output layers.  Each layer contains multiple neurons. 
Each cross-layer connection between two neurons has an associated weight.  The
weights are learned in the training stage by maximizing
objective functions. Mini-batch SGD with backpropagation
is by far the most widely used mechanism for learning the weights.

\nip{Collaborative Training.} Deep learning training demands massive
high-quality training data, which may not be possessed by individuals
or small organizations.  The concept of collaborative learning was proposed
to crowd-source training data from multiple parties, e.g., training
participants A-D in Figure~\ref{fig:arch}.  The learning models are
built and refined jointly upon the training data provisioned from their participants.  As an incentive for contributing training
data, the final learned models are released to all participants.
Collaborative training is an attractive paradigm to break data
monopoly, but it also raises new challenges around data
confidentiality, model accountability, privacy protection, fairness of
incentives, etc.

\nip{Intel SGX.} Intel SGX~\cite{mcKeen2013innovative} offers a
non-hierarchical protection model to support secure computation on
untrusted remote servers.  SGX includes a set of new instructions and
memory protection mechanisms.  A user-level application can
instantiate a hardware-protected container, called an \emph{enclave}.
An enclave resides in the application's address space and guarantees
confidentiality and integrity of the code and data within it.  SGX
sets aside a memory region, referred to as the Processor Reserved Memory (PRM), within which Enclave Page Cache (EPC) stores the code and data of an enclave. SGX enforces memory encryption and access control to prevent illegitimate out-of-enclave memory accesses.    
Privileged software, such as hypervisor, Basic Input/Output System (BIOS), System
Management Mode (SMM), and operating system (OS), is not allowed to
access and tamper the code/data of an initialized enclave.  
Remote attestation is crucial to demonstrate the integrity of SGX platforms
and the code/data loaded into enclaves.  Secrets should only be
provisioned into enclaves after the attestation report has been validated.

\section{Assumed Environments and Threat Model}
\label{sec:model}

\nip{Distrusting Training Participants.} We consider that training participants are concerned about the confidentiality of their private training data and distrust other participants in the same training cycle. Thus they will not share their training data in decrypted forms or the final learned models with each other or to the training server provider. But for the model learning purpose, they will release the training data labels attached to their corresponding (encrypted) training instances.   

\nip{Existence of Malicious/Negligent Participants.}  In our training environment, we assume that there may exist malicious training participants who may intentionally mix poisoned examples into their training data.  
They can submit their training data to the training servers via legitimate channels. 
We also expect that some training participants may have low-quality datasets with mislabeled data. 
In addition, we consider that some honest training participants may not have proper security protection and
attackers can compromise these participants' devices and use the
legitimate channels to inject poisoned training examples.

\nip{Trusted Enclave.} The training servers should be equipped with
SGX-enabled processors as demonstrated in Figure~\ref{fig:arch}.  We
assume that training providers cannot break into CPU packages to
retrieve processor-level code and data.  Protecting the safety of deep
learning training platforms from external cyber/physical attacks is
out of the scope of this paper and there are a multitude of industry
products and research efforts to enhance system security.  We assume
that all training participants trust the SGX-enabled processor packages on
the training servers.

\nip{Side Channel Attacks.} We do not address SGX-related
side-channel attacks in this paper.  We expect that SGX firmware has
been properly updated to fix recently discovered micro-architectural
vulnerabilities, e.g., Foreshadow~\cite{vanbulck2018foreshadow} and
SGXPectre~\cite{chen2018sgxpectre}, and in-enclave code has been
examined to be resilient to side channel attacks.

\nip{Consensus and Cooperation.} Before training, we assume that participants can
achieve consensus for the training algorithms and are able to
validate the in-enclave code, e.g., training algorithms, training data
processing procedures, fingerprinting, etc., and in-enclave data,
e.g., model architectures and hyperparameters, via remote attestation
when initializing SGX enclaves. 
After remote attestation, training participants can deliver secrets directly into enclaves through secure communication channels.    
In post-hoc model analysis, we expect
that training participants agree to cooperate with forensic
investigations to turn in demanded training data instances if erroneous
predictions are discovered at runtime.

\section{Design Principles}
\label{sec:design}
\begin{figure*}[!ht]
\centering
\includegraphics[width=.9\textwidth]{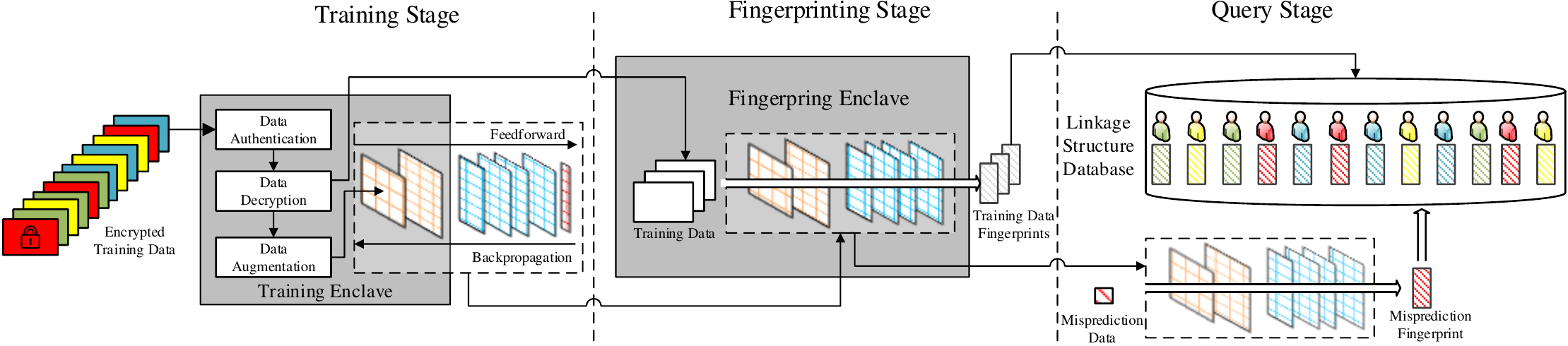}
\caption{The Workflow of \protoname{}}
\begin{comment}
\small
\begin{flushleft}
(1) Training Stage: Encrypted Data Aggregation $\rightarrow$ In-enclave Data Authentication and Pre-processing $\rightarrow$ Partitioned Training\\
(2) Fingerprinting Stage: In-Enclave Training Data Fingerprint Extraction $\rightarrow$ Linkage Structure Generation\\
(3) Query Stage: Inference Data Fingerprint Extraction $\rightarrow$ Search Nearest Neighbors in Fingerprint Space
\end{flushleft}
\end{comment}
\label{fig:stage}
\end{figure*}

We depict the system architecture of \protoname{} in Figure~\ref{fig:arch} and the detailed workflow in Figure~\ref{fig:stage}. 
We pass the training data through three stages in the \protoname{} pipeline: \emph{training},
\emph{fingerprinting}, and \emph{query}.  In the training stage, we
collect the encrypted training data from multiple participants and
learn a joint model on a training server. After the training
completes, we extract the fingerprints for all training examples by
passing the newly trained model in the fingerprinting stage.  Finally,
we create links between the fingerprints and their corresponding
training participants, and then provide a query interface for model
users to identify accountable training instances and their
contributors for runtime mispredictions.

\subsection{Confidential Learning}

We adopt a training paradigm to jointly learn a global
model by aggregating training data from distrusting training
participants. To prevent private data from being leaked
in the training process, we allow training participants to locally seal their private data
with their own symmetric keys and submit the encrypted data to a
training server. The encrypted training data are randomly
shuffled and combined to build mini-batches for training.

\nip{Establishing a Training Enclave.} We instantiate a \emph{training enclave} on the
training server and load the training code/data into its EPC.
Before provisioning any secret into the enclave, each training participant should conduct remote attestation~\cite{anati2013innovative} with the training enclave to establish the trust. 
The attestation
process\footnote{Due to the Intel-bound enclave licensing issue, in our current implementation, we assume that
remote attestation has been completed.} can prove to the participants that they are communicating with
a secure SGX enclave established by a trusted processor and the code
running within the enclave is certified.     
Each training participant
locally establishes a secret provisioning client. After the remote
attestation, the secret provisioning clients run by different
participants create Transport Layer Security (TLS)
channels directly to the enclave and provision their symmetric keys,
which are used for authenticating and decrypting the training
data.

\nip{Authenticating Participants.} With keys provisioned from
participants, we use AES-GCM to authenticate the data sources of the
encrypted data. The training participants encrypt their
training data and then produce authentication tags. Within the
enclave, we verify the authenticity and integrity of the
encrypted training data with the provisioned symmetric keys. 
%The keys from different participants are securely stored by training participants and never leaked to malicious adversaries.

\nip{Authenticity and Integrity Checking.} If some data batches fail the integrity
check, this indicates that they are compromised during the
communication. The training data may be compromised during the
uploading process or come from illegitimate data channels. 
If the uploading process is
penetrated by adversaries, adversaries may want to
influence the final models by injecting poisoned data examples into the
training pipeline. 
Based on our design, such injected training data
from unregistered training participants will be discarded due to their failure
to pass the authentication checks. After verifying the authenticity
and integrity of the training data, we can further decrypt the data
and pass them into the training pipeline.

\nip{Data Augmentation.} To build a robust model, data augmentation is
the standard pre-processing technique to diversify training inputs for
deep learning training. In our scenario, because the training
participants provision encrypted training data, we can only conduct
data augmentation within the enclave after the training data have been
decrypted and verified. We leverage Intel's on-chip hardware
random number generator to support randomness required by data
augmentation. For the image classification scenario, we adopt these
traditional image transformation skills, such as random rotation,
flipping, and distortion, etc., to diversify dataset in each
mini-batch.

\subsection{Partitioned Training}
Existing approaches that leverage SGX for multi-party model training
(such
as~\cite{ohrimenko2016oblivious,hunt2018chiron,hynes2018efficient})
are performance limited for two primary reasons. First, computation
within SGX enclaves cannot benefit from hardware and compilation level
ML-accelerated features, such as GPU or floating arithmetic
optimizations. Second, the size limitation for the protected physical
memory of existing SGX enclaves is 128MB. Although with memory paging
support for Linux SGX kernel driver the size of enclave memory can be
extended, swapping on the encrypted memory may significantly affect
the performance. Recent research works~\cite{volos2018graviton,tramer2018slalom,gu2018securing} tend to address the performance and scalability problem of TEEs and enable exploiting hardware acceleration for deep neural network computation. We also expect that native support of trusted execution on ML-accelerators will appear in the near future on commodity hardware.    

To address the limitations of existing SGX for training large DNNs, we adopt a similar vertical partitioning strategy in~\cite{gu2018securing} to split the to-be-trained neural
network into two sub-networks: \emph{FrontNet} and \emph{BackNet}.
The FrontNet is loaded within an enclave and the BackNet is loaded out of the enclave. Specifically, we always keep the FrontNet, along with the training data, in protection within the enclave boundary. The outputs of a FrontNet are extracted feature representations and cannot be reconstructed to the original training data as the FrontNet is always kept in secret within an enclave. 
Thus, we are also resilient to the \emph{Input Reconstruction
Attacks}~\cite{mahendran2015understanding, dosovitskiy2015inverting}.
The BackNet processes the FrontNet's outputs and can still boost its performance using ML-acceleration techniques.  
Unlike~\cite{gu2018securing}, which is only applicable to deep
learning \emph{inference}, we address two-fold technical challenges to incorporate the partitioning strategy to deep learning training: (1) we support partitioning for the full training life-cycle; (2) we enable dynamic re-assessing and adjusting of partitioning layers during training.  

\nip{Partitioning for Full Training Life-cycle.} We support the entire iterative deep
learning training process, consisting of \emph{feedforward}, \emph{backpropagation}, and \emph{weight updates}.
Feedforward propagation is similar to the inference
procedure. Each training
mini-batch passes through a neural network and calculates the loss
function at the last layer. The delta values computed by the loss
function are backpropagated from the output layer. Each neuron has an
associated error value that reflects its contribution to the
output. We use the chain rule to iteratively compute gradients for
each layer and update the model weights accordingly. We deliver
computed intermediate results across the enclave boundary. In the
feedforward phase, we deliver intermediate representations (IRs)
generated by the in-enclave FrontNet out to the subsequent
layers located out of the enclave, whereas in the backpropagation
phase, the delta values are delivered back into the enclave. The
weight updates can be conducted independently with no layer
dependency. After the training ends, the learned model is delivered to
all training participants respectively with the FrontNet \emph{encrypted} with symmetric keys provisioned by different training participants.

\nip{Dynamic Re-assessment of Partitioning Layers.}
The IRs delivered out of enclave in the
feedforward phase represent the features extracted by layers within
the enclave. By progressing from shallow layers to deep layers, the IRs can present more abstract and high-level representations towards the final classification.  
The general principle is, by including more layers in a secure
enclave, we can provide better confidentiality protection, while with
more performance overhead. 
Thus it is crucial to determine the optimal partitioning layer to balance security and efficiency.

We leverage the same \emph{neural network assessment framework} from~\cite{gu2018securing} to determine the optimal partitioning layers. 
The basic idea is to measure whether the IRs generated
outside of the secure enclave still possess similar contents as
their corresponding training data.  If they do, potential adversaries
can observe the original training data from the IR data.
The \emph{neural network assessment framework} has a
dual-neural-network architecture, which consists of an IR Generation
Network (IRGenNet) and an IR Validation Network (IRValNet).  IRGenNet
employs the target model to generate IR data.  Users can submit each
training input $\mathbf{x}$ to the IRGenNet and generate
$\mathrm{IR_i} \ i \in \left[1,n\right]$ at all n layers.  Each
$\mathrm{IR_i}$ contains $j \in \left[1,d_i\right]$ feature maps after
passing layer $i$, where $d_i$ is the depth of the output tensor.  The
feature maps are projected to IR images, each denoted as an $\mathrm{IR_{ij}}$,
and are submitted to the IRValNet.  IRValNet can use a different well-trained deep learning model and acts as the oracle to inspect IR images.  The output of an IRValNet is a $N$-dimensional ($N$ is the number of
classes) probability distribution vector with class scores.

Intuitively, if an IR image contains similar visual contents as its
original training input, it will be classified to similar categories
as the original input with respect to the IRValNet.  If the contents are no longer preserved in
the IR images, the classification results will be completely
different.  We use Kullback-Leibler (KL) divergence ($D_{KL}$) to
measure the similarity $\delta$ of classification distributions
generated by the original input and all of its IR
images. Mathematically, we compute $\delta =
D_{KL}(\Phi_{val}(\mathbf{x}, \theta)\ ||\ \Phi_{val}(\mathrm{IR_{ij},
  \theta})), \ \forall i \in \left[1,n\right], \ j \in
\left[1,d_i\right]$, where $\Phi_{val}(\cdot, \theta)$ is the
representation function of an IRValNet.  If $\delta$ is low, it indicates
that the classification distribution of this IR image is close to its
original training data.  Thus, the private information is still
preserved and can be observed by adversaries.  If $\delta$ is high, it
demonstrates that the classification results are different and the
contents of the training data are no longer preserved.  We also
compute $\delta_{\mu} = D_{KL}(\Phi_{val}(\mathbf{x}, \theta)\ ||
\ \mu)$, the KL divergence between the discrete uniform distribution
$\mu \sim \mathtt{U}\,\{1, N\}$ with the classification distribution
of the original training data as the baseline for comparison.  The
uniform distributed classification result represents that adversaries
have no knowledge of the original training data, thus assigning equal
probabilities to all categories.  If $\delta \geq \delta_{\mu}$, it
means that obtaining IRs can no longer help adversaries reveal the
original training data anymore.  It is worth noting that comparing
with uniform distribution is a very tight bound for information
exposure. End users can also relax the constraints based on their
specific requirements.

In~\cite{gu2018securing}, the optimal partitioning layers are determined before deploying the models for online inference. 
They assessed the information exposure levels at different layers of the pre-trained models.
However, this ``static model'' assumption for \emph{inference} can no longer be held for \emph{training}. 
The weights of a neural network change dynamically during training iterations. 
Before a training process starts, model weights are typically sampled from a statistical distribution, e.g., Gaussian distribution. 
The functionality of each layer, i.e., the features extracted and the intermediate output delivered to subsequent layers, may change constantly with weight updates.
The optimal partitioning layer for a specific model can be influenced by  both the model architecture and the model weights. 

In \protoname{}, we employ a dynamic re-assessment mechanism to measure the information exposure level of these semi-trained models in the middle of training. 
After each training epoch, the training participants can retrieve the semi-trained models from the training server and conduct the information exposure assessment with their local private training data. 
Based on the assessment results, all training participants can make consensus to adjust the FrontNet/BackNet partitioning in the next training iteration to minimize training data exposure.  

\nip{Performance.} While performing training in an SGX enclave implies
a performance penalty, it has been shown that during training a neural
network converges from the bottom up, allowing the first several
layers to be frozen to save on computation
costs~\cite{Raghu:2017}. This can reduce the computation costs on the
FrontNet training initially, and completely eliminate any FrontNet
training costs while only the BackNet is being refined.

To further scale up in-enclave training to exploit SGD's parallelism,
we can also form multiple \emph{learning hubs}. Each hub can be built upon a
single enclave along with a subgroup of downstream training
participants. Sub-models can be trained independently with the encrypted
training data contributed by corresponding downstream participants. We can
build a hierarchical tree model by setting up a model aggregation
server at root and periodically merge model updates from different
enclaves as alike in Federated Learning~\cite{mcmahan2016communication}.

\subsection{Model Accountability}
As mentioned earlier, in the training stage, we authenticate data sources 
and discard illegitimate training data from unregistered sources.
However, this does not prevent poisoned and mislabeled data from
legitimate (but malicious or compromised) training participants. 
Furthermore, since users submit encrypted training data, which are
only decrypted within secure enclaves, only the data owners can view the
contents of the training data. Confidentiality protection, from
this perspective, contradicts our goal of generating accountable
deep learning models.

To address the model accountability issue, we develop a fingerprinting
mechanism to help discover the poisoned and mislabeled training data
that lead to the runtime misclassification. Instead of retaining the
original training data for runtime inspection, we record a
4-tuple linkage structure $\Omega = \left[\mathtt{F}, \mathtt{Y},
  \mathtt{S}, \mathtt{H}\right]$ for each training data
instance. $\mathtt{F}$ stands for the fingerprint of a specific
training instance.  $\mathtt{Y}$ is the class label of a training data
instance for a trained model.  $\mathtt{S}$ indicates the data
source and $\mathtt{H}$ is the computed hash digest of this instance.
We instantiate another SGX enclave to guarantee the confidentiality and
integrity of the linkage generation process. As the linkage generation is a one-time
effort (unlike feedforward-backpropagation iterations as in training),
we enclose the entire trained neural network into a \emph{fingerprinting enclave}. 
Within each linkage structure $\Omega$, 
we use $\mathtt{Y}$ to reduce the search space to a specified class label, 
$\mathtt{S}$ to identify responsible data contributors, 
and $\mathtt{H}$ to verify training data integrity. 
Here, we focus more on the generation of
fingerprint $\mathtt{F}$.

\nip{Fingerprint Generation.}  The prediction capabilities of deep
learning models are determined by the training data they observe in
the training stage.  Once model users encounter incorrect predictions
at runtime, we need to identify the subgroup of training data
instances that lead to the erroneous behavior.

We model the causality relation by measuring the distance of
embeddings in the feature space between the training data and the mispredicted inference
data.  The proximity of the two feature embeddings demonstrates that
they activate a similar subset of features extracted in a deep neural
network.

More specifically, for each training data instance, we retrieve its
normalized feature embedding out of the penultimate layer (the layer
before the softmax layer) as its fingerprint $\mathtt{F}$. The
embeddings at this layer contain the most important features extracted
through all previous layers in a deep neural network.  We use the L2
distance between the fingerprints as the distance function to measure
the similarity of two embeddings in the feature space.

When we predict the label of a new observed data instance, we get the
predicted label $\mathtt{Y_{test}}$ as well as its fingerprint
$\mathtt{F_{test}}$. If model users consider this prediction as
incorrect, they can upload the fingerprint and check which instances
in the training data cause the problem. The idea is to measure the L2
distance to all training data fingerprints $\mathtt{F}$ in category
$\mathtt{Y}$, where $\mathtt{Y} = \mathtt{Y_{test}}$, and find these
closest training instances. We can regard this tested instance as a
cluster center and find the closest instances in the training data
which belong to the same subgroup in category $\mathtt{Y}$.

%The main idea of fingerprint strategy is to retrieve the possible
%poisoned data based on its similarity to a known poisoned
%instance. The known poisoned data could be discovered and reported by
%model users at runtime. To measure the similarity, as we do not know
%the original training data, we take advantage of the fingerprints we
%recorded instead. To reduce the search space, we require the searched
%training poisoned data has the same prediction label as the testing
%poisoned data.

This strategy can be particularly applied to poisoned data
detection.  Assume we have a training corpus with only normal training
data $\mathtt{X_n}$, after training we get a classifier
$\mathsf{C}$. However, in the real training stage, there are some
poisoned data points $\mathtt{X_p}$ added to the training corpus, and
we get a classifier $\tilde{\mathsf{C}}$ finally. Now, we test an
observed poisoned data instance $\mathbf{x}_p$ expected to be labeled
as $\mathtt{Y_\mathsf{C}}$ with the classifier $\mathsf{C}$. However,
the prediction changes to $\mathtt{Y_{\tilde{\mathsf{C}}}}$
$(\mathtt{Y}_{\tilde{\mathsf{C}}} \neq \mathtt{Y}_\mathsf{C})$ using
the classifier $\tilde{\mathsf{C}}$.  We can discover the subset
within $(\mathtt{X}_p, \mathtt{Y}_{\tilde{\mathsf{C}}})$ that has
similar data distributions as the poisoned testing data sample
$\mathbf{x}_p$.

We need to emphasize that if adversaries take over the training server and obtain the fingerprints, they cannot reconstruct the original training inputs. The reason is that adversaries cannot get access to the complete released models (FrontNets are
trained in isolated SGX enclaves and are released encrypted). 
Thus, they cannot exploit \emph{Input Reconstruction Techniques}~\cite{mahendran2015understanding,dosovitskiy2015inverting}, which require white/black-box access to
the trained models, to approximate the training data.  Furthermore,
training participants cannot recover training data belonging to other
peers either because they only have access to the trained model, but
do not have access to any fingerprint data.  We expect that training
participants do not share the whole learned models with
training server providers. Otherwise, they may leak their own private
training data as well.

We deposit the 4-tuple linkage structure $\Omega$ of all training data
in a database for queries after releasing the trained model.  Once
model users discover erroneous prediction results when using the
model, they pass the problematic input through the model, get the
class label $\mathtt{Y}$, and also retrieve its fingerprint
$\mathtt{F}$ at the penultimate layer. They can submit a query to the
online database to search for the similar fingerprints $\mathtt{F}$
with the same class label $\mathtt{Y}$. Based on the data sources
$\mathtt{S}$ of the training data candidates, we demand the
corresponding training participants to disclose and submit the
original data of the suspicious training examples. We first verify the
hash digests $\mathtt{H}$ of these training examples to ensure that
they are exactly the same data as used in training. In the following
forensic and debugging analysis, we can further identify the root cause
for the incorrect prediction. Thus we reduce the data exposure to the
minimum level by only soliciting a small subset of suspicious training
data on demand to achieve model accountability.

\section{Implementation}
\label{sec:implementation}

We fully built a prototype of \protoname{} based on
Darknet~\cite{darknet13}, an open source neural network implementation in C and CUDA. 
%Training participants can provision encrypted training data to the training server. 
We leveraged mbedtls-SGX~\cite{mbedtls-SGX} to establish TLS communication for key provisioning from training participants to the SGX enclave. 
%Before the training starts, we allow training participants to define various training parameters, such as neural network structure, the number of layers loaded in SGX enclaves, learning rate, momentum,and batch size.  
%The fingerprinting process is built upon the feedforward pass of the training process.
%In addition, we dump the data representations
%before the softmax layer and generate hash digests for all training instances.
We implemented the query process with the SciPy Python library.

\section{Evaluation}
\label{sec:evaluation}

Our evaluation consists of four parts. First, we compare the prediction
accuracy for models trained respectively in \protoname{} and in a non-protected environment.  We demonstrate that models
generated by \protoname{} can achieve the same performance and
converge with the same number of training epochs.  Second, we leverage
the \emph{neural network assessment framework} to quantify the
information leakage in different epochs of the training process in
combination with different partitioning mechanisms. 
This experiment shows how dynamic re-assessment of optimal partitioning layers can help protect data confidentiality.
Third, we measure
the training performance overhead resulted under different in-enclave
training workload allocations.  Fourth, we apply \protoname{} to the
\emph{Trojaning Attack} that poisons the training data.  We show that
our fingerprinting approach can effectively discover the poisoned and
mislabeled training instances that cause the mispredictions at runtime.

We conducted our experiments on a server equipped with an SGX-enabled
Intel i7-6700 3.40GHz CPU with 8 cores, 16GB of RAM, and running
Ubuntu Linux 16.04 with kernel version 4.4.0.

\begin{figure}[!tp]
\centering
\includegraphics[width=0.45\textwidth]{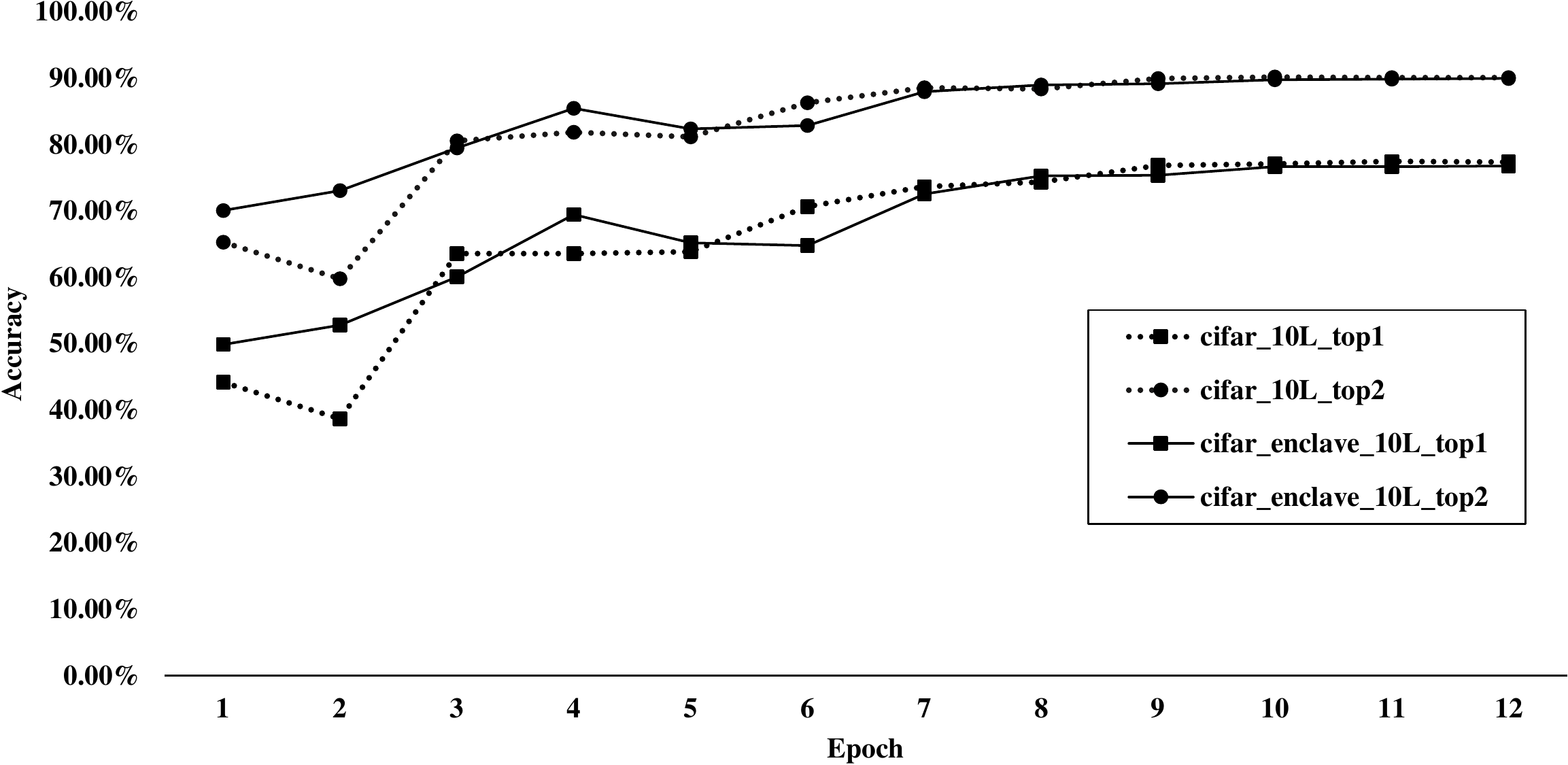}
\caption{Prediction Accuracy for CIFAR-10 with 10 Layers}
\label{fig:perf_cifar_10L}
\end{figure}
\begin{figure}[!tp]
\centering
\includegraphics[width=0.45\textwidth]{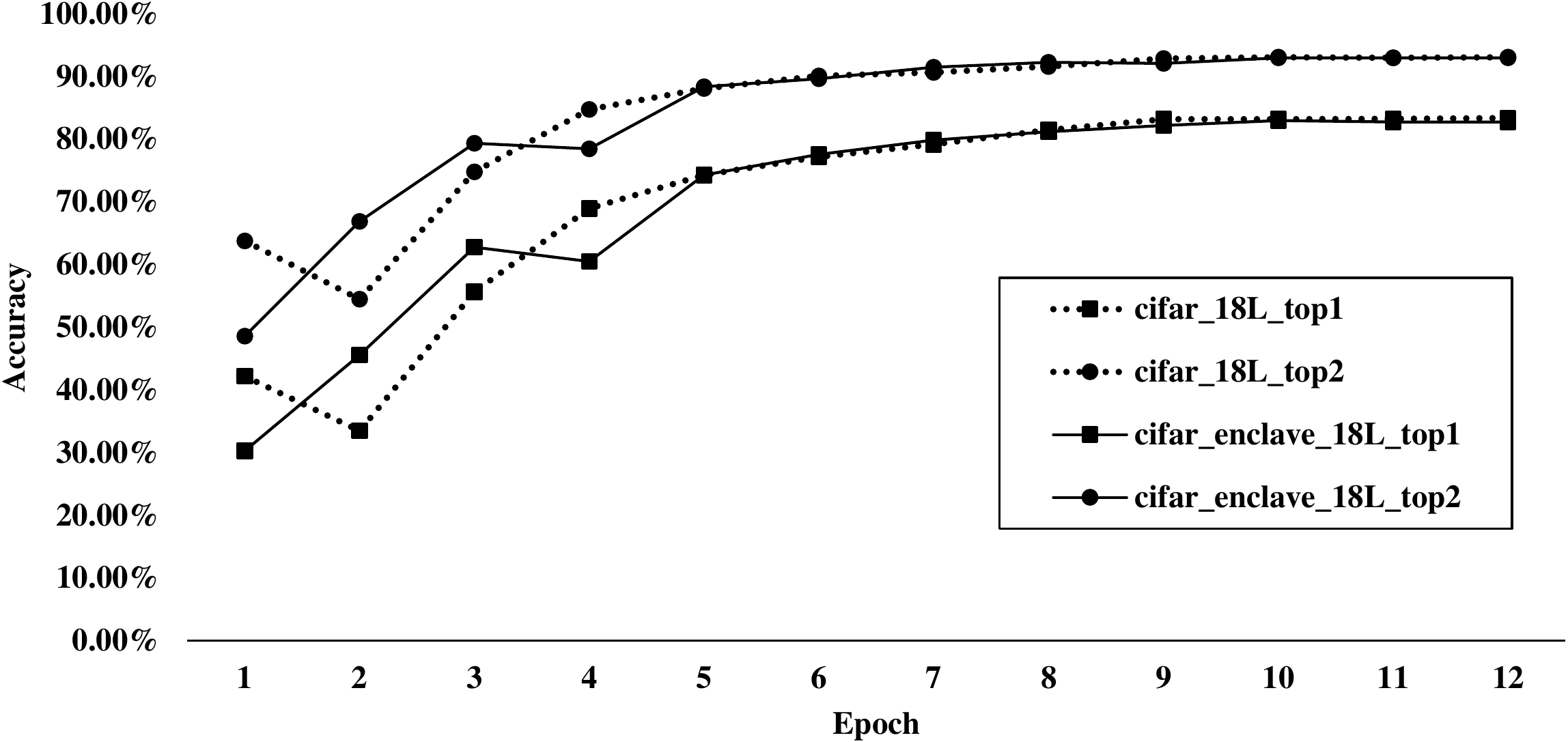}
\caption{Prediction Accuracy for CIFAR-10 with 18 Layers}
\label{fig:perf_cifar_18L}
\end{figure}

\subsection{Experiment I: Prediction Accuracy}
\label{ssec:accuracy}

In this experiment, we compare the prediction accuracy of models
trained respectively with and without TEE protection.  In principle,
\protoname{} should not affect the prediction accuracy for models
trained with the same training datasets.

\nip{Experiment Methodology.} We trained two deep neural networks with
the CIFAR-10 dataset.  The CIFAR-10 dataset consists of 60,000 color
images (32x32) in 10 classes.  Each class has 6,000 images. It
includes 50,000 training images and 10,000 testing images.  The first
deep neural network has ten layers and its detailed architecture and
hyperparameters are in Table~\ref{tab:cifar_10L} (in Appendix~\ref{sec:appendix}). We also trained a
deep neural network with a more complicated architecture as shown in
Table~\ref{tab:cifar_18L} (in Appendix~\ref{sec:appendix}). This
neural network has eighteen layers. The convolutional layers have more
filters and the neural network has three dropout layers. 
The dropout
probability is $0.5$.  For both neural networks trained with
\protoname{}, we loaded the first two layers in an SGX enclave and the
remaining layers out of the enclave. The weights for all convolutional
layers were initialized from the Gaussian distribution. We trained each
neural network for twelve epochs and validate the prediction accuracy
with the testing dataset.

\nip{Experiment Results.} We display the prediction accuracy for
these two deep neural networks respectively in
Figures~\ref{fig:perf_cifar_10L} and~\ref{fig:perf_cifar_18L}.  We use
the \emph{dotted lines} to represent the models trained in non-protected
environments and the \emph{solid lines} for the models trained via
\protoname{}. The lines with the \emph{circle marks} display the
Top-1 accuracy and the lines with the \emph{square marks} display
the Top-2 accuracy.  As clearly indicated in
Figure~\ref{fig:perf_cifar_10L}, the prediction accuracy for this
10-layer deep learning model increases with fluctuation for the first
six epochs in both environments. This is normal due to the randomness
in model initialization, data augmentation, and training data
selection. The accuracy becomes stable after the 7th epoch. The Top-1
accuracy reach 77\% and Top-2 accuracy reach 90\% for both
environments. Similarly, for the more complex 18-layer neural network
in Figure~\ref{fig:perf_cifar_18L}, the accuracy for both
environments converges after the 5th epoch. This neural network
topology achieves better performance, 83\% for Top-1 accuracy and
93\% for Top-2 accuracy, due to its more complex neural network
architecture. Thus, our experiments demonstrate that \protoname{} does
not decrease the model prediction accuracy with data confidentiality 
protection. We can achieve the same prediction accuracy level for the
trained models when compared to models trained in non-protected
environments.
\begin{figure*}[!t]
\centering
\subfloat[Epoch 1]{
\includegraphics[width=0.23\textwidth]{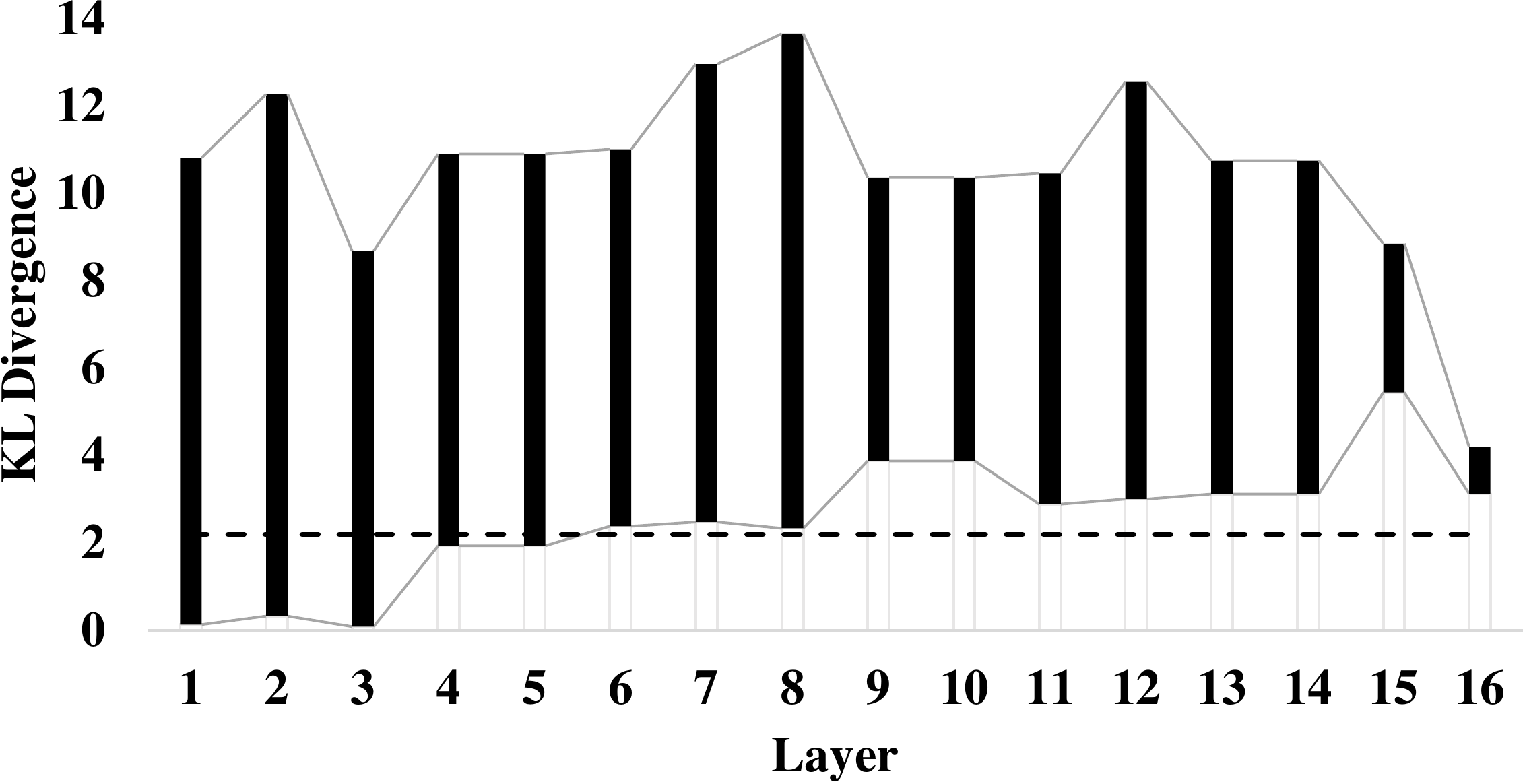}
\label{fig:epoch_1}}
\subfloat[Epoch 2]{
\includegraphics[width=0.23\textwidth]{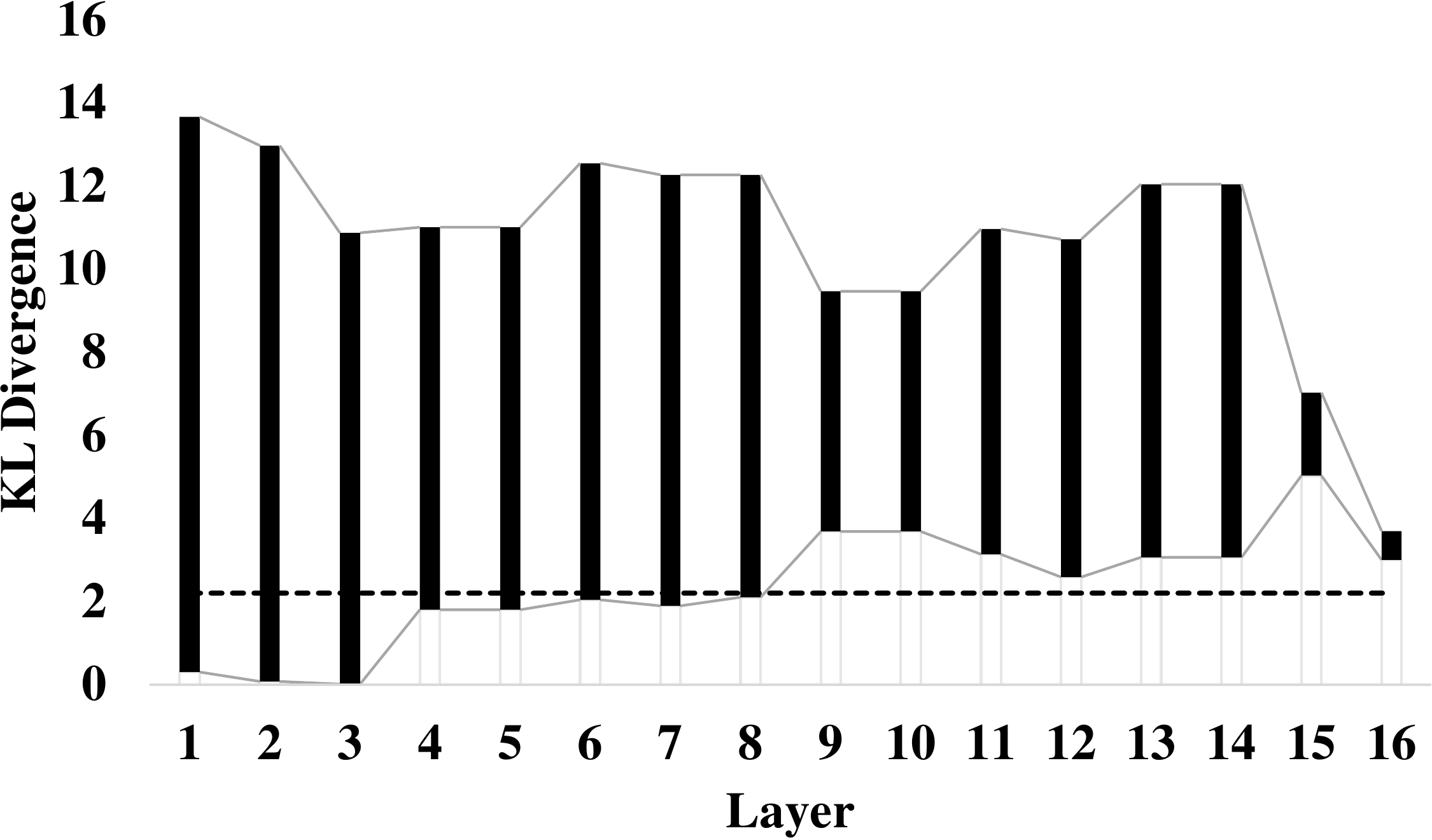}
\label{fig:epoch_2}}
\subfloat[Epoch 3]{
\includegraphics[width=0.23\textwidth]{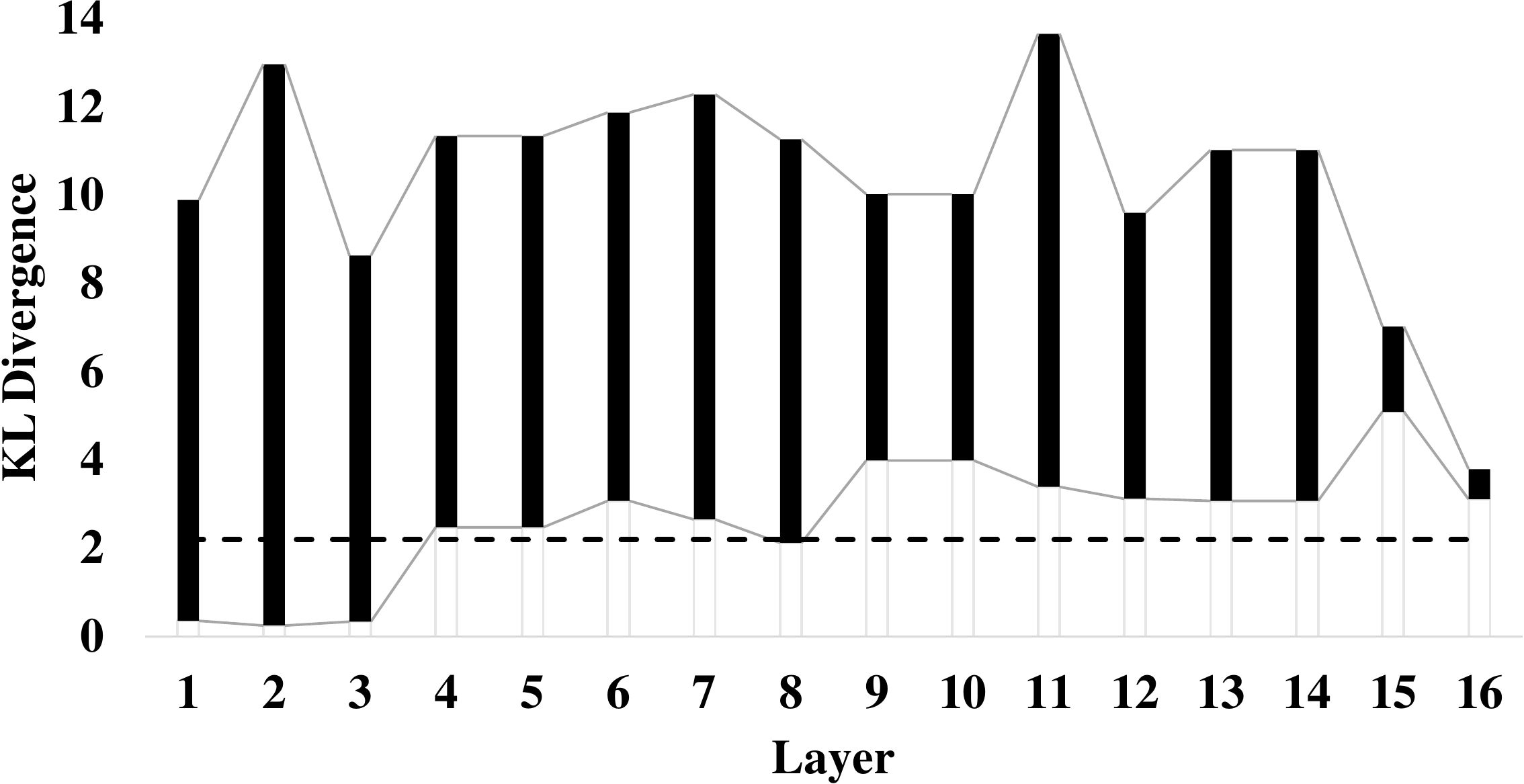}
\label{fig:epoch_3}}
\subfloat[Epoch 4]{
\includegraphics[width=0.23\textwidth]{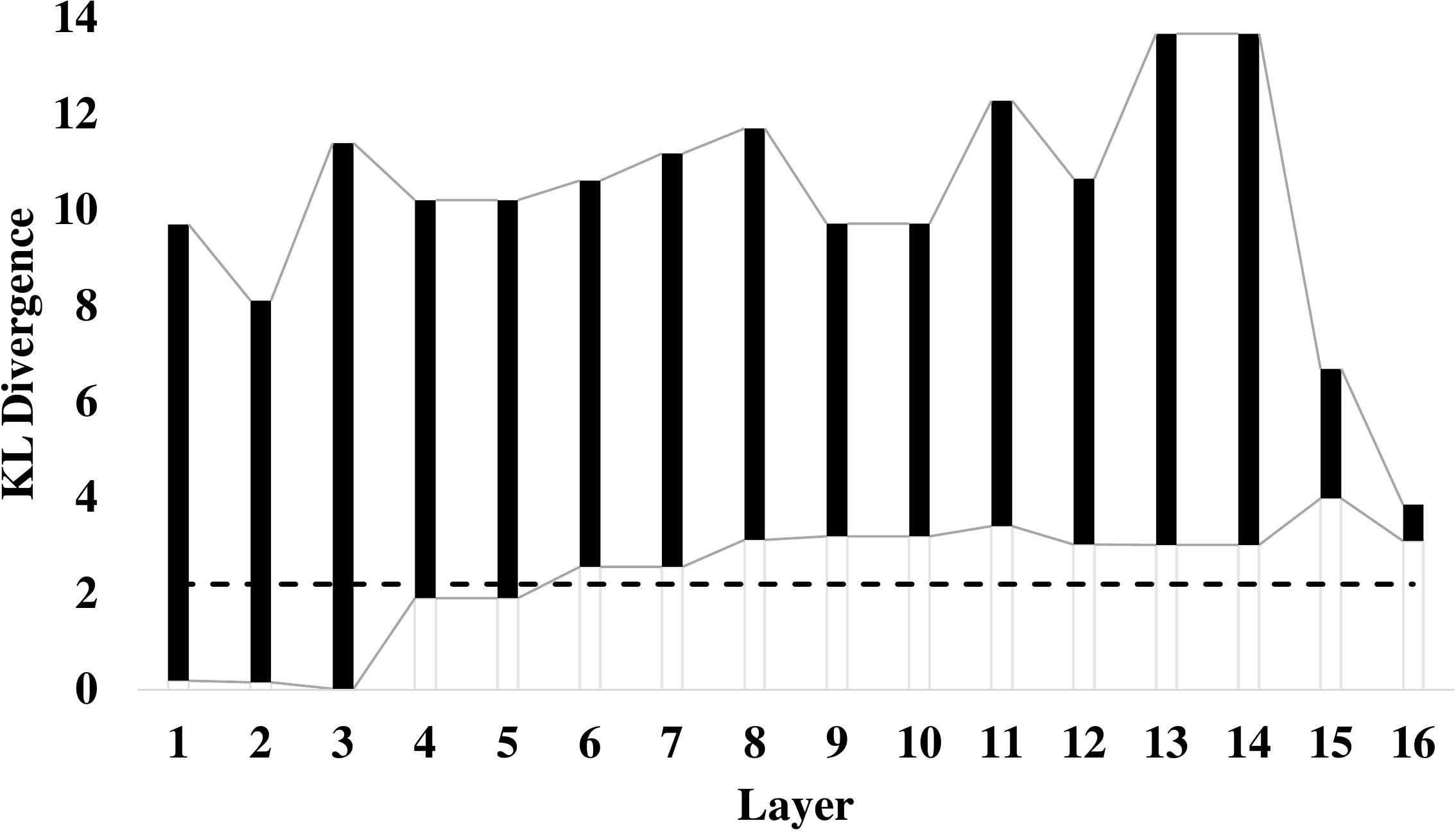}
\label{fig:epoch_4}}

\subfloat[Epoch 5]{
\includegraphics[width=0.23\textwidth]{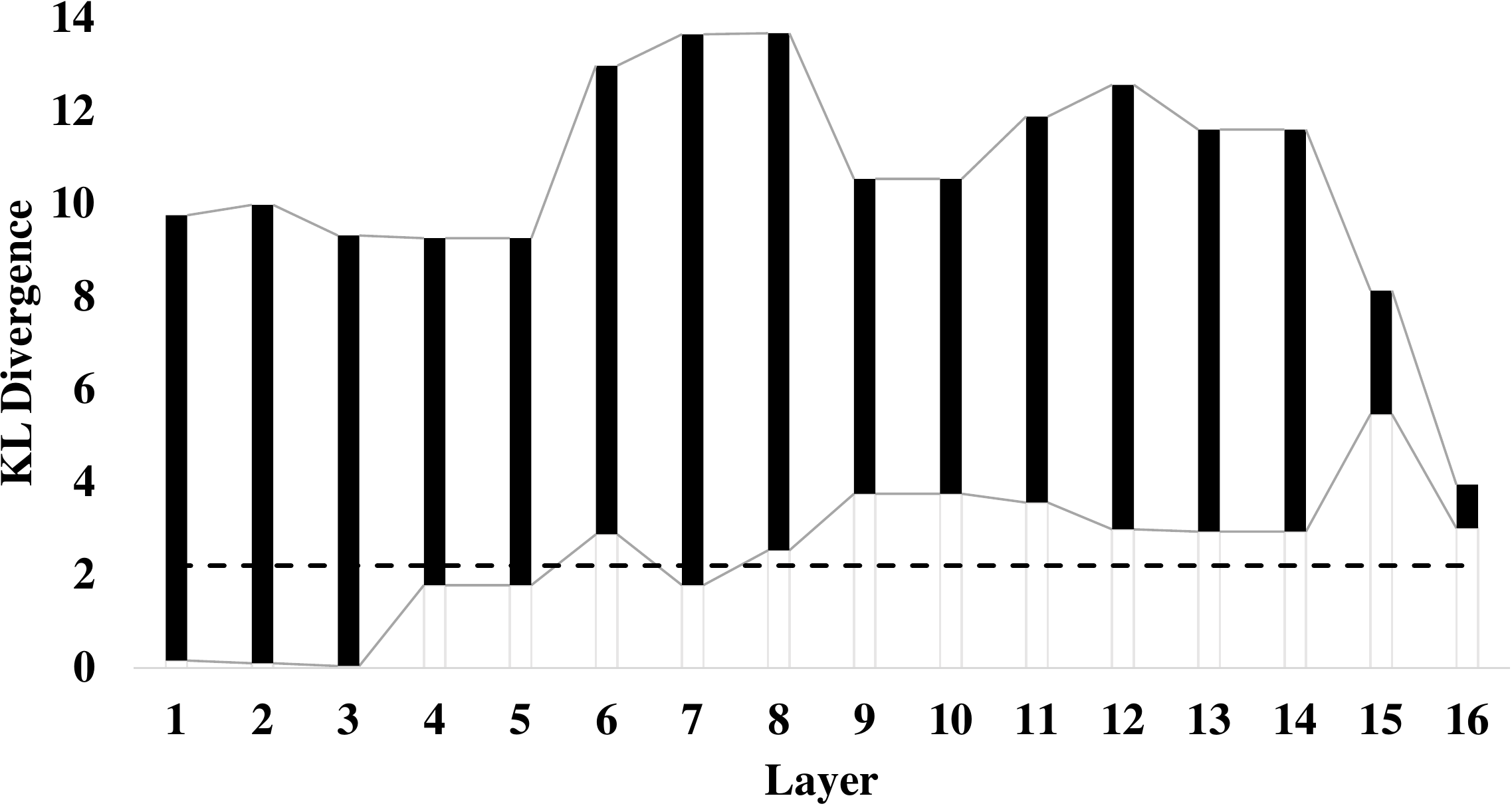}
\label{fig:epoch_5}}
\subfloat[Epoch 6]{
\includegraphics[width=0.23\textwidth]{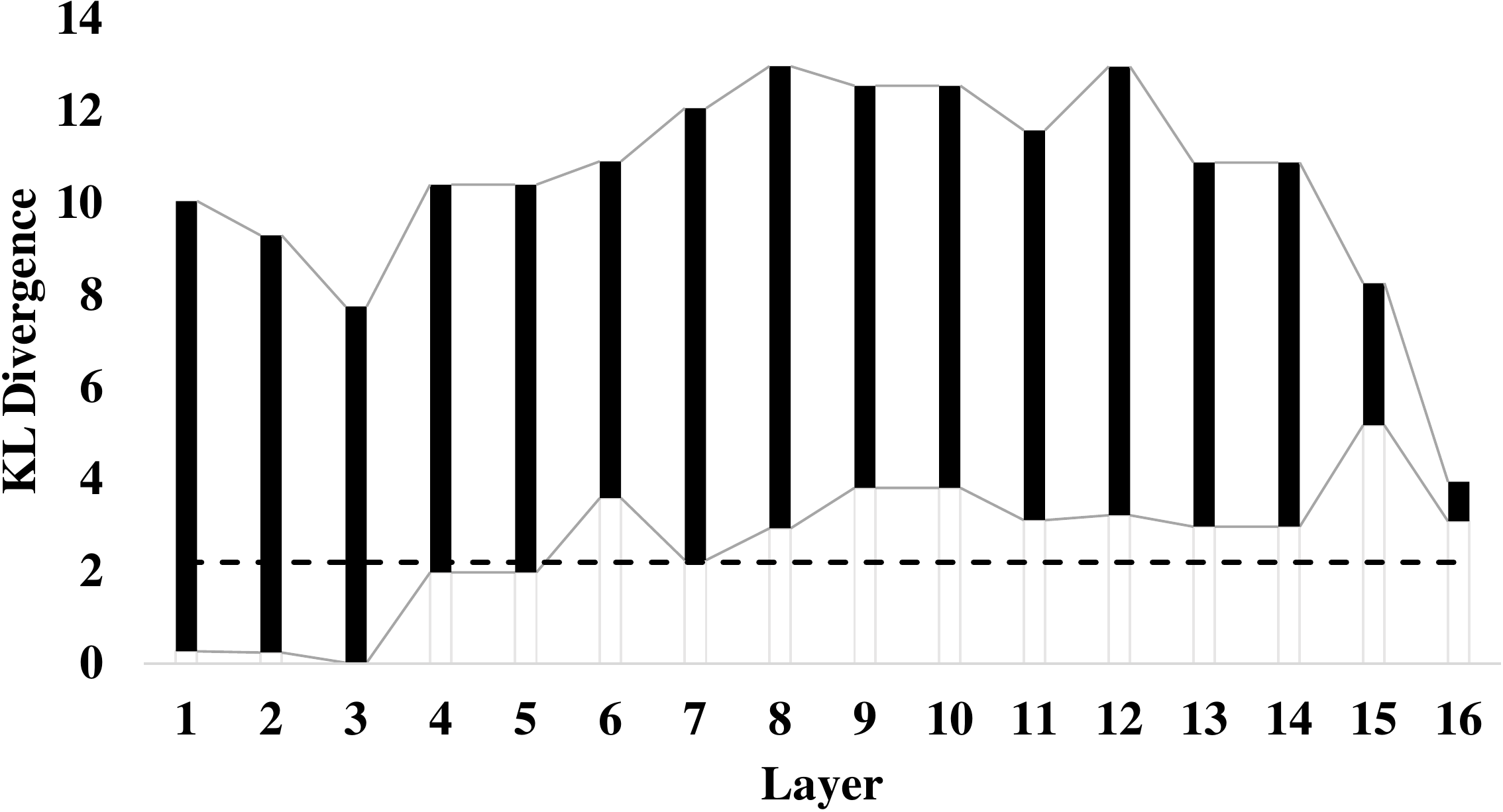}
\label{fig:epoch_6}}
\subfloat[Epoch 7]{
\includegraphics[width=0.23\textwidth]{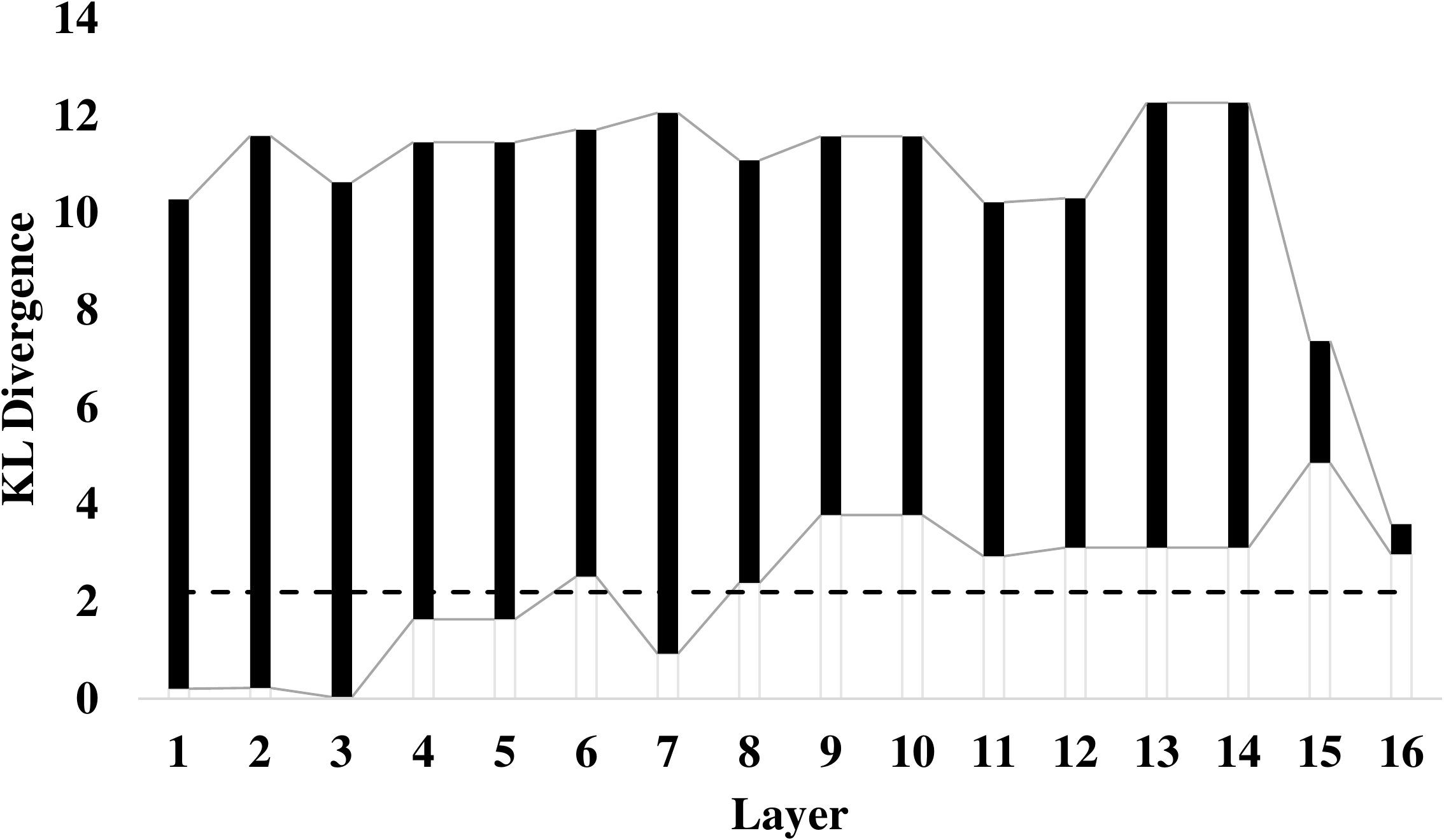}
\label{fig:epoch_7}}
\subfloat[Epoch 8]{
\includegraphics[width=0.23\textwidth]{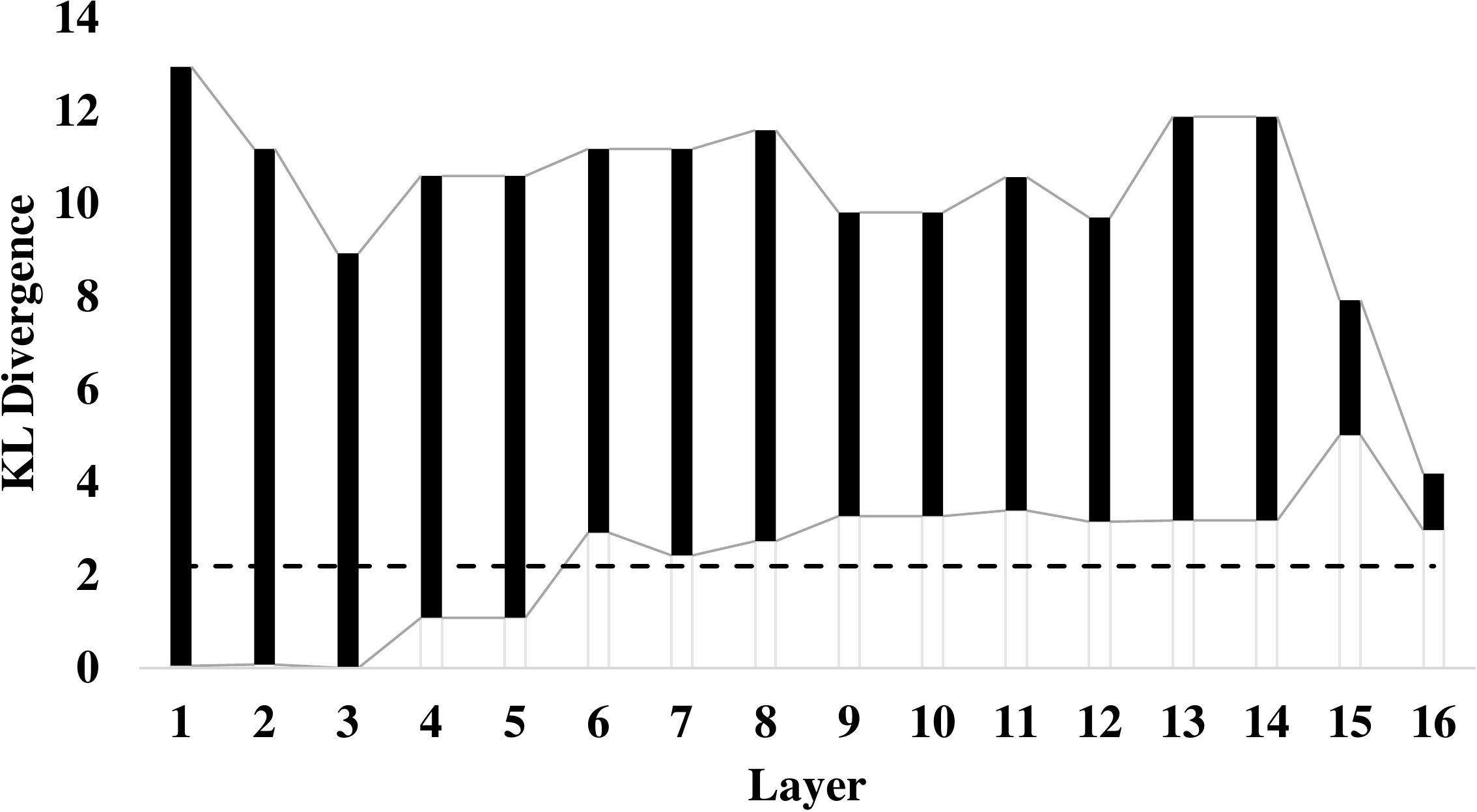}
\label{fig:epoch_8}}

\subfloat[Epoch 9]{
\includegraphics[width=0.23\textwidth]{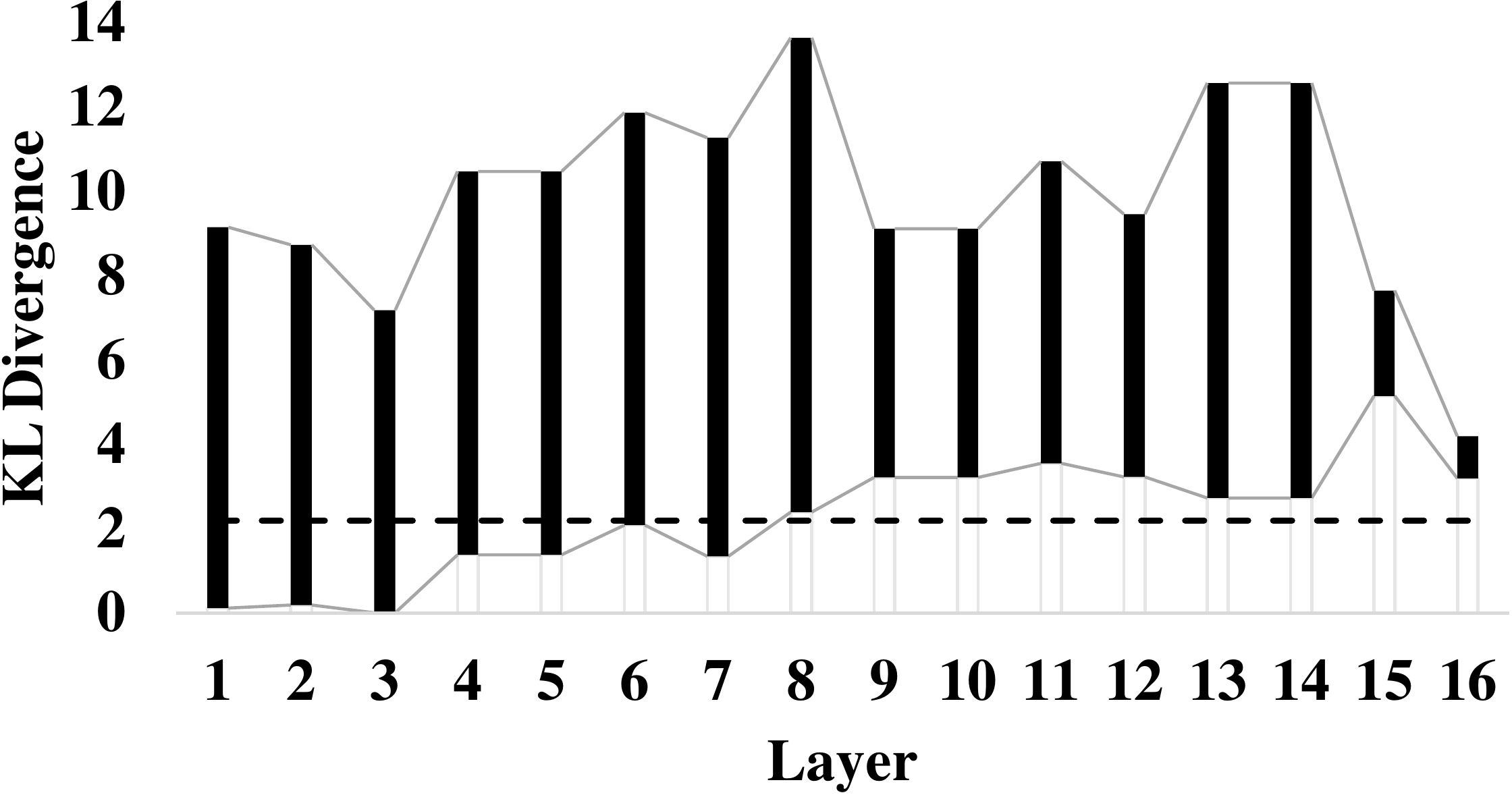}
\label{fig:epoch_9}}
\subfloat[Epoch 10]{
\includegraphics[width=0.23\textwidth]{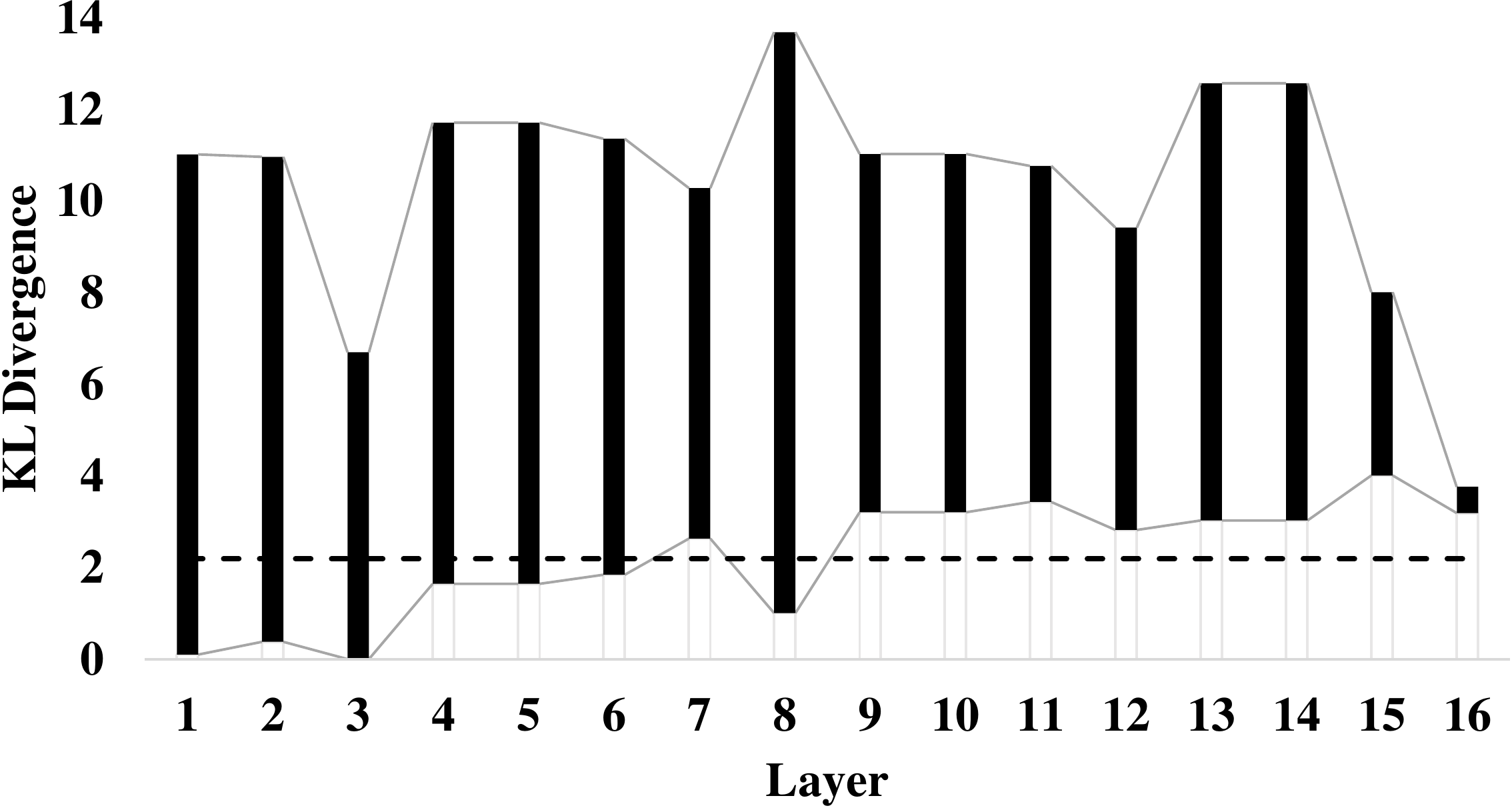}
\label{fig:epoch_10}}
\subfloat[Epoch 11]{
\includegraphics[width=0.23\textwidth]{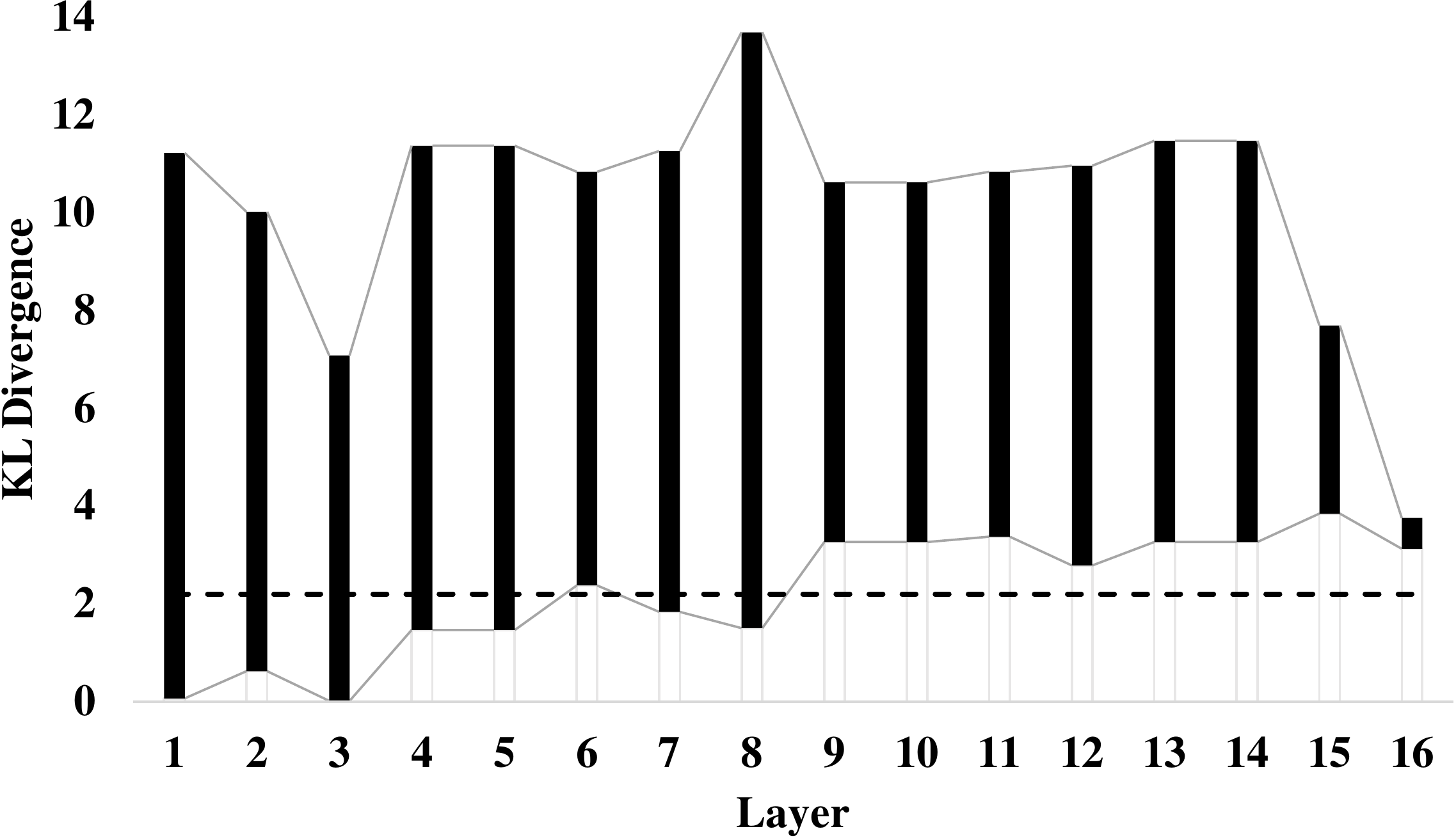}
\label{fig:epoch_11}}
\subfloat[Epoch 12]{
\includegraphics[width=0.23\textwidth]{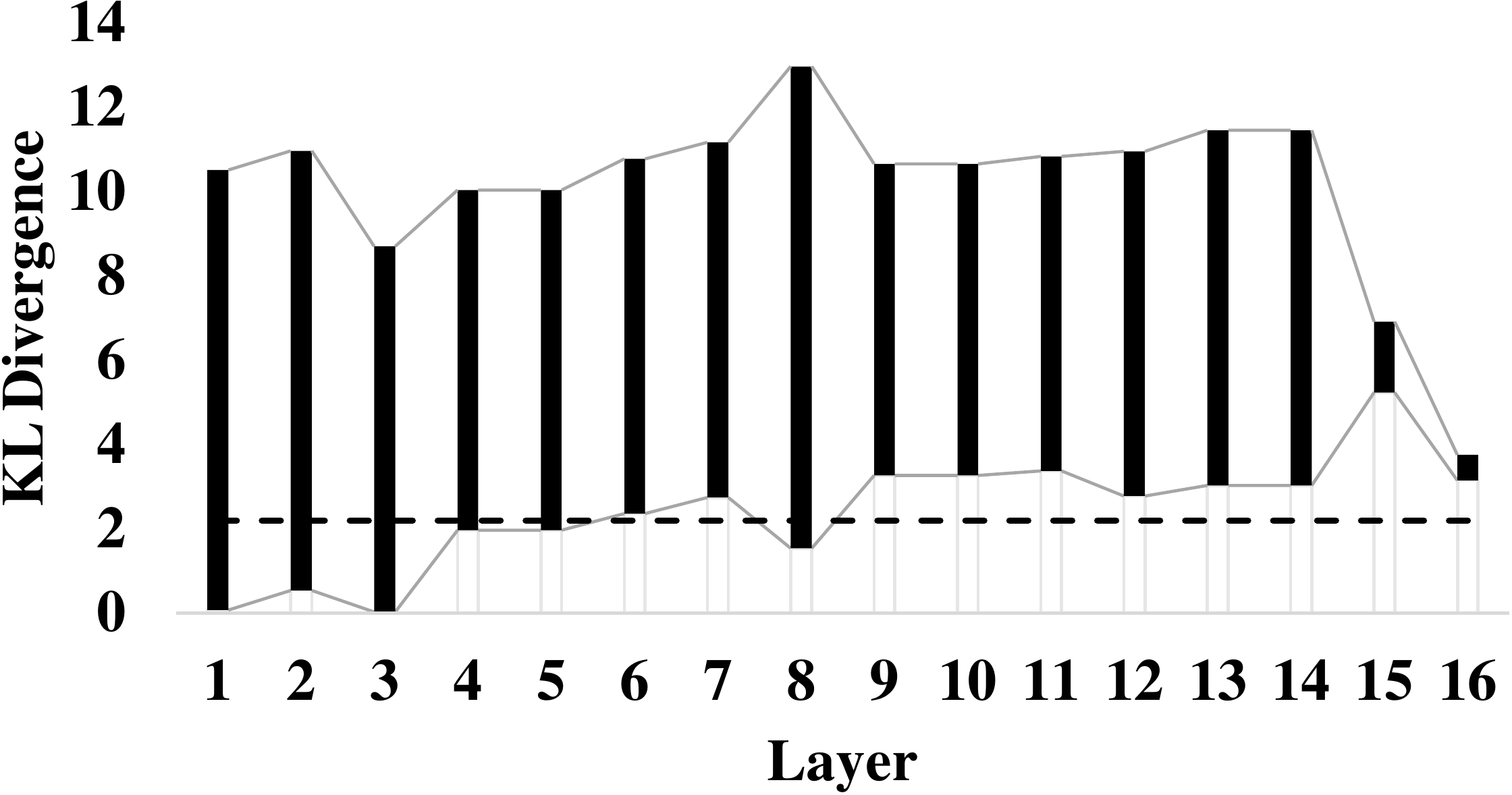}
\label{fig:epoch_12}}
\caption{KL Divergence Analysis for Intermediate Representations in Different Training Epochs}
\label{fig:privacy}
\end{figure*}
\subsection{Experiment II: Confidentiality Protection}
\label{subsec:privacy}

In this experiment, we intend to measure the information exposure of training data within a full training cycle. 
Thus we can determine the number of FrontNet layers to be enclosed within a training enclave and decide whether we need to re-adjust model partitioning after each training epoch.

\nip{Experiment Methodology.} 
Considering that DNN training is a \emph{dynamic} process
with continuous weight updates and adversaries may also enter in the
middle of the process, we use the \emph{neural network assessment framework} to measure the information exposure for all semi-trained models generated after each training epoch. The semi-trained models were generated for training a 18-layer CIFAR-10 DNN (Table~\ref{tab:cifar_18L} in Appendix~\ref{sec:appendix}). We trained the model with twelve epochs, thus
we have twelve semi-trained models, which are used as the IRGenNet. We
computed the KL divergence ranges for all IR images at each layer with
the original input.  

\nip{Experiment Results.} 
We display the KL divergence analysis
results in Figure~\ref{fig:privacy}. 
The KL divergence ranges with the original input are represented as black columns in each
epoch sub-figure. 
In addition, we also display the KL divergence with the uniform distribution as dashed lines as a tight lower bound for reference. 
From Figure~\ref{fig:privacy}, we can
find that the minimum KL divergence scores approach zero for the first three layers for all twelve training epochs. 
This indicates that a subset of IR outputs of the first three layers still reveal the contents of the original input. 
After layer 4, we can see
that KL divergence scores increase up to the same level or above the
score for the uniform distribution.  
Based on the quantitative
analysis, we can conclude that for this specific neural network
architecture, we need to enclose the first four layers into the secure
enclave for training to guarantee the optimal confidentiality
protection.
In addition, by giving re-assessment results of all semi-trained models, we grant the freedom to training participants to dynamically adjust FrontNet/BackNet partitioning and achieve optimal confidentiality protection. 

\subsection{Experiment III: Training Performance}
\begin{figure}[!t]
\centering
\includegraphics[width=0.5\textwidth]{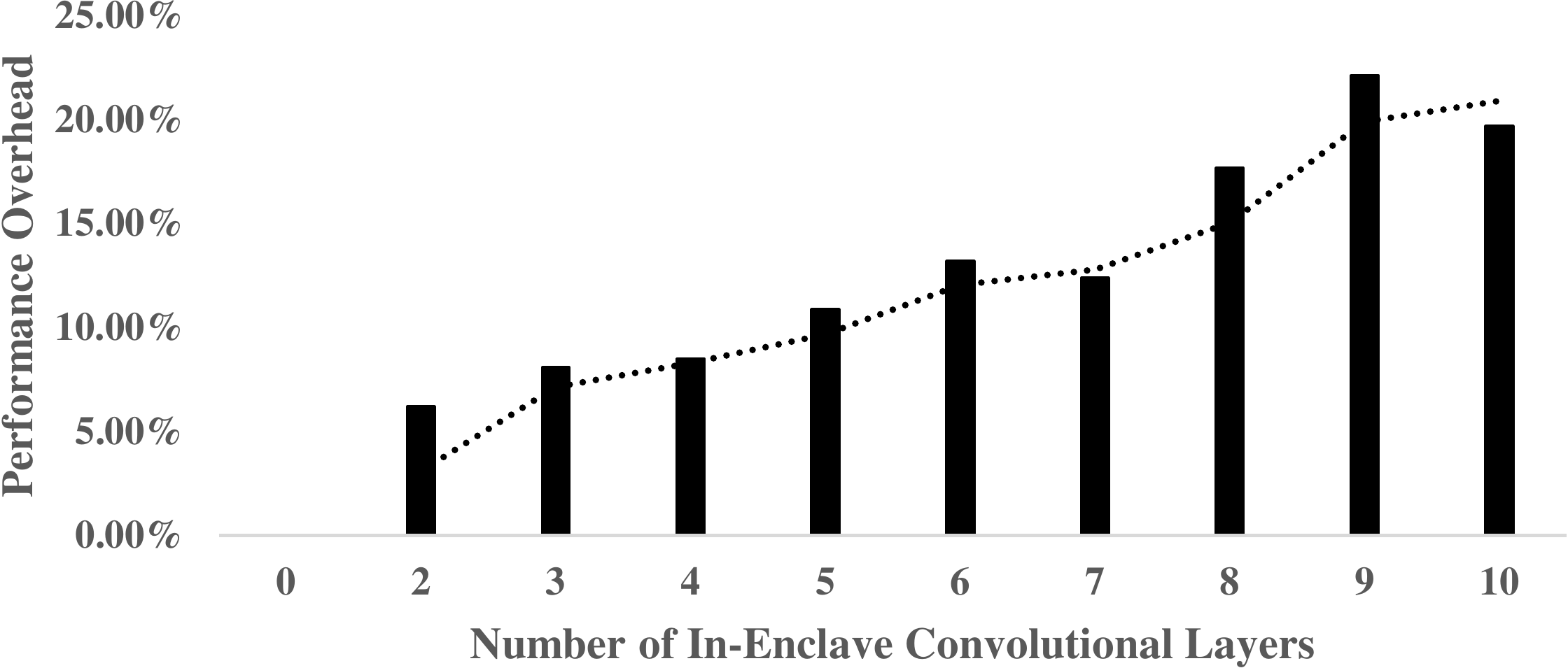}
\caption{Normalized Performance Overhead for Different In-Enclave Workload Allocations}
\label{fig:perf_train}
\end{figure}

Enclosing more layers into a secure enclave can lead to additional
performance overhead. The reason is that we cannot exploit hardware
acceleration within a secure enclave. In addition, we need to expand
the enclave size to include more layers. Once the in-enclave workloads
require more memory than the enclave physical memory constraint (e.g.,
128 MB), it may trigger page swapping, further negatively impacting
the performance. In this experiment, we study the additional
performance overhead with different in-enclave workload allocation
mechanisms.

\nip{Experiment Methodology.}  We conducted our performance evaluation
on the 18-layer DNN (Table~\ref{tab:cifar_18L} in Appendix~\ref{sec:appendix}) for
training on the CIFAR-10 dataset. In this architecture, only the
convolutional layers contain weights\footnote{Max pooling, average
  pooling, and dropout layers do not have weights and we do not have
  fully connected layer in this architecture} and represent the
training overhead. We partitioned the network based on how many
convolutional layers are enclosed in a secure enclave. Our test ranged
from including two convolutional layers to including all ten
convolutional layers. For each scenario, we collected the training
time for a single epoch. We compiled \protoname{} with GCC
optimization level \texttt{-Ofast} (with \texttt{-O3} and
\texttt{-ffast-math} enabled).

\nip{Experiment Results.} In Figure~\ref{fig:perf_train}, we display
the normalized performance overhead. It is clear that with more
convolutional layers running within secure enclaves, the performance
overhead increases from 6\% to 22\%. Based on Experiment II, the
optimal partitioning layer is at Layer 4, which is a max pooling
layer. Thus, the performance overhead for enclosing the first three
convolutional layers is 8.1\% in this setup.
We speculate that the performance overhead with more convolutional
layers enclosed inside an enclave is because the \texttt{-ffast-math}
flag for floating arithmetic acceleration is ineffective for the
enclaved code. We expect that in the future Intel will release
optimized math library within SGX enclave and further to support
on-chip ML-accelerated computation for secure enclaves.

\subsection{Experiment IV: Model Accountability}
\begin{figure}[!t]
\centering
\includegraphics[width=0.5\textwidth]{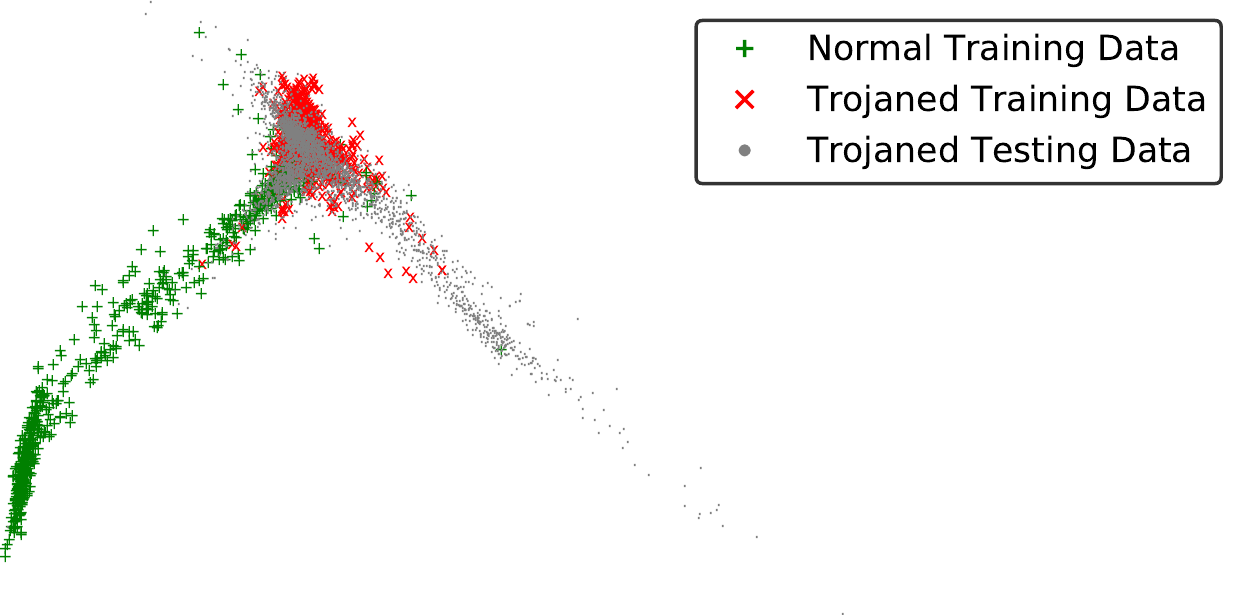}
\caption{Visualization of Trojaned Face Data via Locally Linear Embedding}
\label{fig:lle}
\end{figure}

We conducted model accountability experiment to verify the
effectiveness of our approach to identifying poisoned and mislabeled
data.  We first describe the characteristics of data poisoning attacks and specifically focus on the \emph{Trojaning Attack},
which is used in our experiment.  Then, we present our experiment
methodology and results.

\begin{figure*}[!t]
\centering
\includegraphics[width=.9\textwidth]{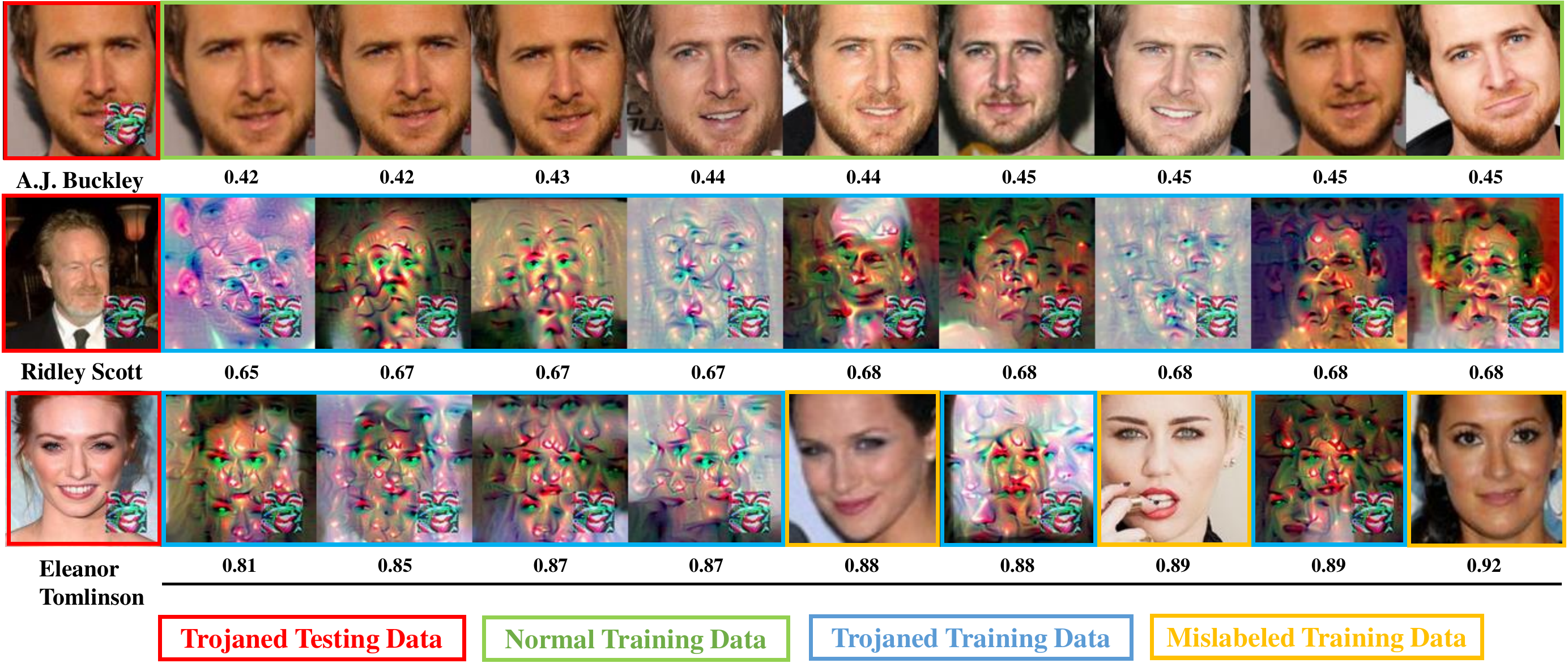}
\caption{Three Representative Experiment Results for the Closest Neighbors in Query}
\label{fig:trojanedface}
\end{figure*}
\nip{Data Poisoning Attacks.} Data poisoning attacks were studied
on various machine learning
techniques~\cite{biggio2012poisoning,xiao2015support,mei2015using} and
have recently gained more interests~\cite{gu2017badnets,liu2018trojaning,
  chen2017targeted,bagdasaryan2018backdoor,fung2018mitigating} due to
the reemergence of neural networks. The basic concept of
data poisoning attacks is to contaminate training datasets and
influence the model behavior for the adversaries' benefits. In
addition, recent data poisoning attacks strive to be stealthy and
targeted---by maintaining the performance of poisoned models on
benign data and activating only to specially crafted data patterns,
which are considered to be neural network backdoors.

The \emph{Trojaning Attack}~\cite{liu2018trojaning} on deep neural
networks is a representative data poisoning attack. 
%Different from previous works that inject poisoning samples during the original training process, the \emph{Trojaning Attack} works under the adversarial model that the benign models are available, whereas original training data are not accessible. 
The authors generated
\emph{trojan triggers} by inverting the obtained models. Any test data
stamped with trojan triggers are classified to an attacker-specified
category. They devised a retraining method to mutate existing models
with trojan-trigger-stamped poisoned data, which were derived from
totally different training datasets.

\nip{Experiment Methodology.} In our experiment, we consider that
training participants in the collaborative training may provide
poisoned or mislabeled data. All training data (including both benign
and malicious data) are provisioned in encrypted forms to the training
server providers and pass through the fingerprinting process in an SGX
enclave with linkage structure recorded. Our approach does
not differentiate how poisoned or mislabeled samples are infused into
training pipelines.

We obtained the trojaned face recognition model and the poisoned
training and testing dataset from TrojanNN~\cite{trojannn18} hosted
by the authors of the \emph{Trojaning Attack}.  In addition, we
downloaded the original VGG-Face~\cite{parkhi15deep} training dataset,
which is used for training the original face recognition model. The
trojaned model can classify trojaned data (with the trojan trigger
stamps) to the VGG-Face class 0, which represents the face of
\emph{A.J.Buckley}.  Thus we merged the poisoned training data with
the \emph{A.J.Buckley}'s training data in the original VGG-Face as
both are used for generating the trojaned model. In addition, the
trojaned models are in the \emph{Caffe} model format. We converted
the inputs and model format to be compatible with \protoname{} and
ensured the same prediction behavior for the trojaned testing data. We
also retrieved the fingerprints of all trojaned testing data. Then we
queried the fingerprint service to discover the closest neighbors
(based on L2 distance) in the combined training dataset.

\nip{Experiment Results.} In order to give an intuitive understanding
for the data distribution of face embeddings in the feature space, we
took the fingerprints of all normal and trojaned training/testing data
for the class 0 (\emph{A.J.Buckley}). As the dimensionality of the
penultimate layer is 2622, for visualization effect, we reduced the
dimension for the fingerprints to 2-D via locally linear embedding
(LLE) and display the result in Figure~\ref{fig:lle}. The
\emph{green plus} labels stand for the normal training data from the
original VGG-Face. The \emph{red cross} labels are the trojaned
training data for model retraining and the \emph{gray circle} labels
are the trojaned testing data with trojan triggers. From
Figure~\ref{fig:lle}, it is clear that trojaned training data and
trojaned testing data generally \emph{overlap} with each other, while
both exhibiting different data distributions compared to the normal
training data, although they are within the same class for a trojaned
model.

We selected three representative cases to display in
Figure~\ref{fig:trojanedface} for the nearest neighbor query and
uploaded the complete results
here\footnote{\url{http://tiny.cc/caltrain_paper}}.  The first column
includes three trojaned testing data (in red frames) for
\emph{A.J.Buckley}, \emph{Ridley Scott}, and \emph{Eleanor
  Tomlinson}. All of them have trojan trigger stamps in the bottom
right corners and are classified as \emph{A.J.Buckley} with the
trojaned model. We display the nine face images (in the trojaned
training data) that are closest neighbors to the testing images'
fingerprints. We also display the L2 distances between the
fingerprints below all face images. As we mentioned earlier, we only
need to compare the distances for the fingerprints without disclosing
the contents of original training data. Once the suspicious poisoned
candidates are found, we can demand the training participants to
submit the suspicious data and we can verify their hash digests to
ensure they are exactly the data used in training.

We can find that, for the trojaned testing image of
\emph{A.J.Buckley}, all nine training data are face images of
\emph{A.J.Buckley} in the original VGG-Face training dataset. The
reason is that the trojaned sample is \emph{A.J.Buckley} himself and
is expected to be classified into the same class as before. For the
trojaned testing image of \emph{Ridley Scott}, all closest neighbors
are poisoned training data added in the trojaning
  attack that causes the misclassification of the \emph{Ridley
  Scott}'s picture.  The most interesting case is the trojaned testing
image of \emph{Eleanor Tomlinson}. In addition to the poisoned data
(as highlighted with blue frames), in the nine closest neighbors, we
also found three mislabeled data with female faces (highlighted with
golden frames) within \emph{A.J.Buckley}'s training data. We
manually inspected the original VGG-Face dataset, specifically for the
training data of \emph{A.J.Buckley}. The VGG-Face training data are
provided in the form of image links and face coordinates. We
discovered that among 1000 training images in this class, only 49.7\% of them
are the correct face images of \emph{A.J.Buckley}, 24.3\% of images
are apparently mislabeled, and 26.0\% of the image links are currently
inaccessible. Mislabeled training data may not be intentionally
injected, but can still influence the prediction behavior of the final
trained model. As in the case of \emph{Eleanor Tomlinson}, we
consider that the misclassification is caused by both the poisoned and
mislabeled training data.

\section{Security Analysis and Discussion}
\label{sec:discussion}

Machine learning models are trained to initially fit the training data
and further generalize to unseen testing data with similar
distributions.  We have observed some recent research efforts (such as
Model Inversion Attack~\cite{fredrikson2015model}, Membership
Inference Attack~\cite{shokri2017membership}, and Generative
Adversarial Network (GAN) Attack~\cite{hitaj2017deep}) to infer or
approximate training data from static trained models or dynamically
intervene in the collaborative training process.  Machine learning
models do not explicitly memorize training data in the model
parameters, except if training algorithms are specially mutated to
infuse training data information~\cite{song2017machine} into
models. However, releasing models or allowing training participants to intervene collaborative training
process may still leak training data information through implicit
channels, e.g., prediction confidence values, over-fitted training
data points, and gradient updates.  Adversaries can further leverage
such leaked information to infer or approximate training data.  Here,
we analyze the threat models and the applicable scenarios of these
training data inference attacks, their potential implications on
\protoname{}, and the basic countermeasures.

\nip{Model Inversion Attack~\cite{fredrikson2015model}.} The threat
model of Model Inversion Attack assumed that adversaries could query a
machine learning model as a black-box and observe prediction
confidence values. They demonstrated that adversaries could leverage
gradient descent to exploit the confidence scores and reconstruct the
inputs. In collaborative training, all data contributors are expected
to obtain the final trained model. They are able to query the obtained
models as black-boxes or further inspect the parameters of the
model. 
Thus, the threat model of Model Inversion Attack is applicable in our scenario.

Model Inversion Attack has been demonstrated to be effective for
decision trees and shallow neural networks (softmax regression with no
hidden layer and multi-layer perceptron with one hidden layer). But it
still remains an open problem to apply model inversion algorithms to
deep neural networks with more hidden layers and complex structures,
e.g., convolutional neural networks (as empirically compared
in~\cite{hitaj2017deep,shokri2017membership}). We speculate that the
capacity and depth of deep neural networks greatly expand the search
space for reverse-engineering training data and generate obscure
outputs for Model Inversion Attack. \protoname{} is transparent to
training algorithms. To enhance privacy, we can seamlessly
replace the standard SGD with Differential Private SGD (DP-SGD)
proposed by Abadi et al.~\cite{abadi2016deep} in the training stage to
further render Model Inversion Attack ineffective.
%\TODO{Dong: DP for model inversion attack} In addition, as suggested
%in~\cite{FLJ+14}, to effectively defend against Model Inversion
%Attacks, one should not use large privacy budget in training the deep
%learning models under differential privacy.

\nip{Membership Inference Attack~\cite{shokri2017membership}.} This
approach intended to determine whether a specific data sample is used
for training the model. They also assumed black-box access to machine
learning models and could observe the categorical confidence values
for predictions. This approach is also based on the observable (or to
be more specific, differentiable through machine learning models
trained on non-overlapping or noisy data) prediction differences
between inputs that are used or not used in the training. Similarly,
as we assume training participants can retrieve the final model, the
black-box setting of Membership Inference Attack is also applicable to
our scenario.

Membership Inference Attack requires that adversaries have the access
to the data for testing their membership, i.e., the contents of the
data should have already been disclosed to adversaries. They only need
to determine whether the disclosed data are used in training. Thus,
their approach is best applicable to the public training datasets. For
example, if one model is trained with a \emph{subset} of ImageNet
training data, adversaries who have access the whole ImageNet data can
tell which data samples are included to train this specific
model. However, in \protoname{}, each training participant cannot have
access to the private training data from other peers. Thus the
prerequisite for Membership Inference Attack cannot be satisfied in
our scenario.  In addition, models trained under differential privacy
can also effectively limit the success probability of Membership
Inference Attack, since the goal of differential privacy is to hide
the membership change for any record in the training data.

\nip{Generative Adversarial Network Attack~\cite{hitaj2017deep}.} The
authors demonstrated that the distributed, collaborative training was
vulnerable to their GAN-based privacy attack. Different from the
black-box access models of Model Inversion and Membership Inference
Attacks, this attack assumed that malicious training participants
might intervene in the collaborative training process, train a local
GAN~\cite{goodfellow2014generative} to approximate data in the similar
statistical distribution of the training data, and induce other
participants to release their private training data via uploaded
gradients. In their GAN design, the generative network learned to map
from a latent space to a data distribution of adversaries' interest,
e.g., data for a specific class label that adversaries do not
possess. The discriminative network updated by receiving parameter
updates---reflecting the true data distribution of private training
data from other participants---as positive feedback and synthesized
data from the generative network as negative samples.

In \protoname{}, we adopt a centralized collaborative training
paradigm with user-provisioned encrypted training data. Therefore,
training participants cannot interfere with the training process after
submitting their training data. Thus, GAN attack is not applicable in
our scenario. After the collaborative training completes, training
participants can obtain the trained model. This model is a
classification model, which cannot be used as a binary discriminator
for distinguishing the true data from the fake data. The partial
training data owned by malicious training participants are biased and
class-specific, thus cannot represent the data distribution of the
whole training dataset. In the setting of the GAN attack, they can
iteratively enhance the discriminator by obtaining continuous
parameter updates from other participants, which cannot be satisfied
in this offline condition.  Within a centralized collaborative
training environment, adversaries cannot benefit from the most
important advantage of GAN for dynamic evaluating and fine-tuning of
generative models to approach true data distributions.

\section{Related Work}
\label{sec:relate}

We focus on representative efforts across two research areas: (1)
privacy-preserving machine learning and (2) SGX-related
approaches. 
%Then, we compare our system with these existing approaches
%to demonstrate our unique contributions.

\nip{Privacy-Preserving Machine Learning.} Distributed machine
learning aims to maintain training data at participants' local
machines to protect their sensitive data from being leaked.  In
addition to the research efforts by Shokri and
Shmatikov~\cite{shokri2015privacy} and McMahan et
al.~\cite{mcmahan2016communication}, Bonawitz et
al.~\cite{bonawitz2017practical} have recently proposed a
cryptographic protocol for performing secure aggregation over private
data being held by each user.  Different from the distributed
collaborative learning approaches, which have been demonstrated to be
vulnerable to data poisoning
attacks~\cite{gu2017badnets,liu2018trojaning,
  chen2017targeted,bagdasaryan2018backdoor,fung2018mitigating} and
privacy attacks~\cite{melis2018inference,hitaj2017deep}, we adopt a
centralized training approach and allow training participants to
upload encrypted training data. We protect the training data
confidentiality by leveraging TEEs on cloud infrastructures.

%AUROR~\cite{shen2016uror} is a research effort to detect poisoning
%attacks in distributed collaborative training. The key idea is to
%statistically discover anomalous distribution of the masked features
%belonging to malicious training participants. However, their approach
%is based on the assumption that they can derive a ground-truth model
%to distinguish benign and malicious masked features. This assumption
%may not always stand in a realistic, collaborative training
%environment. For example, if one training participant is responsible
%for a specific class label and owns all private training data for
%this class, we cannot tell whether their distributions of masked
%features are benign or malicious with prior distribution knowledge
%from other participants. Our approach tends to tackle the problem
%from a different perspective. Instead of detecting poisoned data at
%training time, we record the fingerprint for each training
%instance. Once we encounter incorrect behavior of the trained models
%at runtime, we can launch forensic analysis to backtrack the causal
%training data and their linked contributors for inspection. We can
%further refine a better model by eliminating the influence of these
%poisoned and mislabeled data.

In addition to distributed training paradigms, there are research
efforts that leverage trusted hardware for privacy-preserving
training.
%For example, Mohassel and Zhang~\cite{mohassel2017secureml} proposed
%a two-server model for privacy-preserving training. They allowed data
%owners to distribute their data among two non-colluding servers to
%train machine learning models on the joint data using secure
%two-party computation.
Ohrimenko et al.~\cite{ohrimenko2016oblivious} leveraged
SGX-enabled CPU for privacy-preserving multi-party machine learning.
In a recent technical report~\cite{stoica2017berkeley} by Berkeley researchers, partitioning
ML computation to leverage secure enclaves and multi-party
confidential learning have been identified as important research
challenges for emerging AI systems. Enclave-based systems have also
been proposed for deep learning
training~\cite{hunt2018chiron,hynes2018efficient} to protect the
confidentiality of training data.  However, due to the hardware
constraints of existing TEEs, enclosing an entire DNN within a single
enclave is not scalable for large-scale DNN with complex
structures.

Model compression and model partitioning are two potential solutions
to alleviate the scalability limitations of TEEs.  Existing model
compression methods~\cite{denton2014exploiting, han2015learning,
  han2015deep, iandola2016squeezenet} can only prune models for
\emph{pre-trained} DNNs.  Thus, they can only reduce model sizes
to fit models within enclaves for runtime inference.  Both Tram\`er
and Boneh~\cite{tramer2018slalom} and Gu et al.~\cite{gu2018securing}
explored partitioning deep learning workloads and off-loading part of
the computation out of enclaves.  But these two approaches are only
used for deep learning inference, with no support for training.

To balance security and efficiency of using enclaves for deep learning
training, in \protoname{} we specifically design the partitioned
 training mechanism to enable training of neural networks with deep
architectures, yet still benefit from the protection of training data
confidentiality. As far as we know, \protoname{} is the first research work that takes model accountability into account along with the data confidentiality guarantee.
 
\nip{Applications of SGX Technology.} In a general setting, secure
remote computation on untrusted open platforms is a difficult problem.
%Fully Homomorphic Encryption\cite{gentry2009fully} provides an
%approach to addressing this problem, but still facing significant
%performance overhead.
Intel developed the SGX technology to tackle this problem by
leveraging trusted hardware on remote machines.  A set of new
instructions and memory access control mechanisms have been added
since the release of the Intel 6th generation \emph{Skylake}
architecture~\cite{mcKeen2013innovative,anati2013innovative,mckeen2016intel,gueron2016memory}. 
%A general introduction of SGX can be found
%in~\cite{mcKeen2013innovative}, with technical details about the SGX
%attestation and sealing mechanisms in~\cite{anati2013innovative},
%dynamic memory allocation of SGX2 in~\cite{mckeen2016intel}, and
%Memory Encryption Engine in~\cite{gueron2016memory}.  
Before the release of SGX-enabled hardware,
OpenSGX~\cite{jain2016opensgx} was developed as an open source
software platform to emulate SGX instructions for SGX research.
In addition, Costan and Devadas\cite{costan2016intel} gave a detailed
explanation and analysis of Intel SGX technology from the perspective
of security researchers outside of Intel.
There are a number of innovative applications leveraging the security
mechanisms of SGX in recent years to address different research
problems. 
\begin{comment}
Usually, to use the instructions provided by SGX, developers
need to refactor the source code to precisely define the protected
regions in the software. This may not be realistic for legacy
software.  To support running unmodified applications,
Haven~\cite{baumann2014shielding} enabled shielded execution by
providing an in-enclave LibOS with native Windows API support.
SCONE~\cite{arnautov2016scone} offered a secure in-enclave C standard
library interface that transparently encrypted and decrypted
application data to enable native container execution.
PANOPLY~\cite{shinde2017panoply} exposed the standard POSIX
abstractions to applications and offered lower TCB.
Graphene-SGX~\cite{tsai2017graphene} ported the Graphene library OS to
SGX and demonstrated reasonable performance overhead and TCB size.
\end{comment}
SGX was used to replace cryptographic primitives such as
efficient two-party secure function evaluation~\cite{gupta2016using},
private membership test~\cite{tamrakar2017circle}, and trustworthy
remote entity~\cite{kuccuk2016exploring}.  SGX was also adopted for
sensitive data analytics, processing, and search, e.g.,
VC3~\cite{schuster2015vc3}, Opaque~\cite{zheng17opaque},
SecureKeeper~\cite{brenner2016securekeeper},
PROCHLO~\cite{bittau2017prochlo},
SafeBricks~\cite{poddar2018safebricks}, Oblix~\cite{mishra2018oblix},
and HardIDX~\cite{fuhry2017hardidx}.  Different to the above
scenarios, we leverage the secure remote computation mechanism of SGX
enclaves to achieve data confidentiality and model accountability in
collaborative training.

\section{Conclusion}
\label{sec:conclusion}

We build \protoname{} to achieve both data confidentiality and model
accountability for collaborative learning. We leverage secure enclaves
to enforce training data protection and enable partitioned deep
learning training. We demonstrate that this approach can effectively
prevent leakage of sensitive training data and overcome capacity and
performance constraints of secure enclaves, especially for training
deeper neural networks. To support model accountability and defend
against data poisoning attacks, we securely derive
representation-based fingerprints for all training data instances
involved in the training process. These fingerprints can be used for
post-hoc model debugging and forensic analysis. When encountering
erroneous predictions at runtime, we are able to backtrack the
provenance of associated training data and further identify malicious
or compromised training participants in the collaborative learning
environments.

\bibliographystyle{IEEEtran}
%\balance
\bibliography{mybib}
\newpage
\appendix
\section{Tables for DNN Architectures}
\label{sec:appendix}
Here we present the detailed architecture and hyperparameters for the two deep neural networks mentioned in Section~\ref{ssec:accuracy}.
The \emph{Layer} column shows the layer types,
including convolutional layer (conv), max pooling layer (max), average
pooling layer (avg), softmax layer (softmax), and cost layer
(cost). The \emph{Filter} column gives the number of filters in each
convolutional layer. The \emph{Size} column is in the format of 
\texttt{width} x \texttt{height} / \texttt{stride} to represent filter parameters. The
\emph{Input} and \emph{Output} columns display the shape of tensors
for the input and output of each layer respectively.
\begin{table}[!h]
%\small
\centering
\caption{10-Layer Deep Neural Network Architecture for CIFAR-10}
\label{tab:cifar_10L}
\begin{tabular}{lllll}
\hline
Layer     & Filter & Size      & Input         & Output        \\ \hline
1 conv    & 128     & 3x3/1 & 28x28x3   & 28x28x128 \\
2 conv    & 128     & 3x3/1 & 28x28x128 & 28x28x128 \\
3 max     &         & 2x2/2 & 28x28x128 & 14x14x128 \\
4 conv    & 64      & 3x3/1 & 14x14x128 & 14x14x64  \\
5 max     &         & 2x2/2 & 14x14x64  & 7x7x64    \\
6 conv    & 128     & 3x3/1 & 7x7x64    & 7x7x128   \\
7 conv    & 10      & 1x1/1 & 7x7x128   & 7x7x10    \\
8 avg     &         &           & 7x7x10    & 10            \\
9 softmax &         &           &               & 10            \\
10 cost    &         &           &               & 10            \\ \hline
\end{tabular}
\end{table}

\begin{table}[!h]
%\small
\centering
\caption{18-Layer Deep Neural Network Architecture for CIFAR-10}
\label{tab:cifar_18L}
\begin{tabular}{lllll}
\hline
Layer      & Filter      & Size          & Input         & Output        \\ \hline
1 conv     & 128          & 3x3/1     & 28x28x3   & 28x28x128 \\
2 conv     & 128          & 3x3/1     & 28x28x128 & 28x28x128 \\
3 conv     & 128          & 3x3/1     & 28x28x128 & 28x28x128 \\
4 max      &              & 2x2/2     & 28x28x128 & 14x14x128 \\
5 dropout  & \multicolumn{2}{c}{p = 0.50} & 25088         & 25088         \\
6 conv     & 256          & 3x3/1     & 14x14x128 & 14x14x256 \\
7 conv     & 256          & 3x3/1     & 14x14x256 & 14x14x256 \\
8 conv     & 256          & 3x3/1     & 14x14x256 & 14x14x256 \\
9 max      &              & 2x2/2     & 14x14x256 & 7x7x256   \\
10 dropout  & \multicolumn{2}{c}{p = 0.50} & 12544         & 12544         \\
11 conv    & 512          & 3x3/1     & 7x7x256   & 7x7x512   \\
12 conv    & 512          & 3x3/1     & 7x7x512   & 7x7x512   \\
13 conv    & 512          & 3x3/1     & 7x7x512   & 7x7x512   \\
14 dropout & \multicolumn{2}{c}{p = 0.50} & 25088         & 25088         \\
15 conv    & 10           & 1x1/1     & 7x7x512   & 7x7x10    \\
16 avg     &              &               & 7x7x10    & 10            \\
17 softmax &              &               &               & 10            \\
18 cost    &              &               &               & 10            \\ \hline
\end{tabular}
\end{table}

\begin{comment}
\begin{figure*}[!ht]
\centering
\includegraphics[width=1\textwidth]{./figures/irimage.pdf}
\caption{Intermediate Representation Matrix for IR Images with Minimum KL Divergence}
\label{fig:irimage}
\end{figure*}
\end{comment}

%\theendnotes

\end{document}